\documentclass[eqsecnum,twocolumn,showpacs,amsmath,amssymb]{revtex4}
\usepackage{bm}
\usepackage{graphicx,subfigure,afterpage}
\usepackage{amsfonts}

\newcommand{\etal}{{\it et al.}}

\newcommand{\vect}[1]{\underline{#1}}
\newcommand{\tens}[1]{\underline{\underline{#1}}}

\newcommand{\vm}{\vect{v}_{\,\rm m}}
\newcommand{\vs}{\vect{v}_{\,\rm s}}

\newcommand{\vrel}{\vect{v}_{\, \rm rel}}
\newcommand{\vv}{\vect{v}}

\newcommand{\ny}{N_y}
\newcommand{\dt}{\Delta t}

\newcommand{\etam}{\eta_{\,\rm m}}
\newcommand{\etas}{\eta_{\,\rm s}}

\newcommand{\Dm}{\tens{D}_{\,\rm m}}
\newcommand{\Ds}{\tens{D}_{\,\rm s}}

\newcommand{\Omm}{\tens{\Omega}_{\,\rm m}}

\newcommand{\nablu}{\vect{\nabla}}

\newcommand{\gdotb}{\bar{\dot{\gamma}}}
\newcommand{\phib}{\bar{\phi}}

\newcommand{\be}{\begin{equation}}
\newcommand{\ee}{\end{equation}}
\newcommand{\bea}{\begin{eqnarray}}
\newcommand{\eea}{\end{eqnarray}}

\newcommand{\gdot}{\dot{\gamma}}

\newcommand{\gae}{\stackrel{>}{\scriptstyle\sim}}

\newcommand{\ie}{{\it i.e.\/}}
\newcommand{\eg}{{\it e.g.\/}}

\newcommand{\versus}{{\it vs.\/}}

\newcommand{\etabar}{\bar{\eta}}
\newcommand{\bw}{\begin{widetext}}
\newcommand{\ew}{\end{widetext}}

\newcommand{\bmini}{\begin{minipage}}
\newcommand{\emini}{\end{minipage}}

\newcommand{\phil}{\phi_\ell} 
\newcommand{\phic}{\bar{\phi}_{\rm c}}
\newcommand{\gdotc}{\bar{\gdot}_{\rm c}}
\newcommand{\model}{d-JS-$\phi$}

\newcommand{\sigmas}{\Sigma_{\rm sel}}
\newcommand{\gdoth}{\gdot_{\rm h}}
\newcommand{\gdotl}{\gdot_\ell}

\begin{document}

\title{Flow phase diagrams for concentration-coupled shear banding}
\author{S. M. Fielding}
\email{physf@irc.leeds.ac.uk} 
\author{P. D. Olmsted}
\email{p.d.olmsted@leeds.ac.uk}
\affiliation{Polymer IRC and
  Department of Physics \& Astronomy, University of Leeds, Leeds LS2
  9JT, United Kingdom} 
\date{\today} 
\begin{abstract}
  After surveying the experimental evidence for concentration coupling
  in the shear banding of wormlike micellar surfactant systems, we
  present flow phase diagrams spanned by shear stress (or strain-rate)
  and concentration, calculated within the two-fluid, non-local
  Johnson-Segalman (\model) model. We also give results for the
  macroscopic flow curves $\Sigma(\gdotb,\phib)$ for a range of
  (average) concentrations $\phib$. For any concentration that is high
  enough to give shear banding, the flow curve shows the usual
  non-analytic kink at the onset of banding, followed by a coexistence
  ``plateau'' that slopes upwards, $d\Sigma/d\gdotb>0$. As the
  concentration is reduced, the width of the coexistence regime
  diminishes and eventually terminates at a non-equilibrium critical
  point $[\Sigma_{\rm c},\phib_{\rm c},\gdotb_{\rm c}]$.  We outline
  the way in which the flow phase diagram can be reconstructed from a
  family of such flow curves, $\Sigma(\gdotb,\phib)$, measured for
  several different values of $\phib$. This reconstruction could be
  used to check new measurements of concentration differences between
  the coexisting bands. Our \model\ model contains two different
  spatial gradient terms that describe the interface between the shear
  bands.  The first is in the viscoelastic constitutive equation, with
  a characteristic (mesh) length $l$. The second is in the
  (generalised) Cahn-Hilliard equation, with the characteristic length
  $\xi$ for equilibrium concentration-fluctuations. We show that the
  phase diagrams (and so also the flow curves) depend on the ratio
  $r\equiv l/\xi$, with loss of unique state selection at $r=0$.  We
  also give results for the full shear-banded profiles, and study the
  divergence of the interfacial width (relative to $l$ and $\xi$) at
  the critical point.

\end{abstract}
\maketitle

\section{Introduction}
\label{sec:intro}

\begin{figure}[h]
\begin{center}
 \includegraphics[scale=1.0]{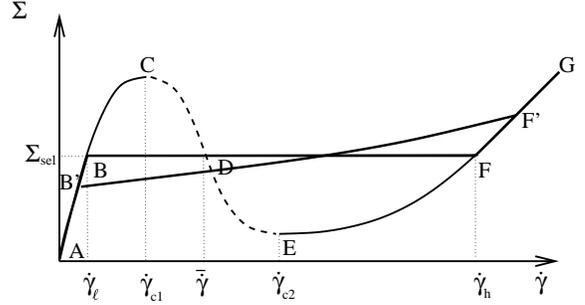}
\caption{Schematic flow curves for  wormlike micelles: the homogeneous
  constitutive curve is ACEG; the steady shear-banded flow curve is BF
  (without concentration coupling in planar shear) or B'F' (with
  concentration coupling, or in a cylindrical Couette device). 
\label{fig:schem} } 
\end{center}
\end{figure}

\begin{figure}[h]
  \includegraphics[scale=0.3]{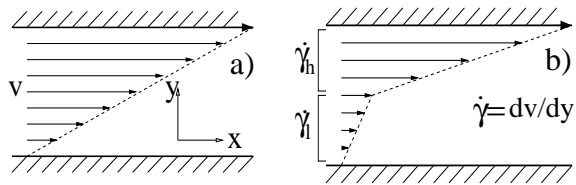}
\caption{(a) Homogeneous shear rate and (b) banded profiles.
\label{fig:picture} } 
\end{figure}

For many complex fluids, the intrinsic constitutive curve of shear
stress $\Sigma$ as a function of shear rate $\gdot$ is non-monotonic,
admitting multiple values of shear rate at common stress. For
semi-dilute wormlike micelles, theory~\cite{cates90,SpenCate94,SCM93}
predicts the form ACEG of Fig.~\ref{fig:schem}. In the range
$\gdot_{\rm c1}<\gdot<\gdot_{\rm c2}$ where the stress is decreasing,
steady homogeneous flow (Fig.~\ref{fig:picture}a) is
unstable~\cite{Yerushalmi70}. For an applied shear rate $\bar{\gdot}$
in this unstable range, Spenley, Cates and McLeish~\cite{SCM93}
proposed that the system separates into high and low shear rate bands
($\gdoth$ and $\gdotl$; Fig.~\ref{fig:picture}b) and that
any change in the applied shear rate then merely adjusts the relative
fraction of the bands, while the stress $\sigmas$ (which is common to
both) remains constant.  The steady state flow curve then has the form
ABFG. Several constitutive models augmented with interfacial gradient
terms have captured this
behaviour~\cite{olmsted99a,lu99,olmstedlu97,spenley96}.


Experimentally, this scenario has been widely observed in semidilute
wormlike micelles~\cite{berret94b,Call+96,grand97}. The steady state
flow curve (which is often attained only after very long
transients~\cite{grand97}) has a well defined, reproducible stress
plateau $\sigmas$.  Coexistence of high and low viscosity bands has
been observed by NMR
spectroscopy~\cite{Call+96,MairCall96,MairCall96c,BritCall97}. Further
evidence comes from small angle neutron scattering
(SANS)~\cite{berret94a,schmitt94,Capp+97,rehage91,berret94b}; and from
flow birefringence (FB)~\cite{Decr+95,Makh+95,DCC97,BPD97}, which
reveals a (quasi) nematic birefringence band coexisting with an
isotropic one (but see~\cite{FisCal01,FisCal00}).

In some systems, the coexistence plateau is not perfectly flat, but
slopes upward slightly with increasing shear rate (B'F' in
Fig.~\ref{fig:schem}). See, for example, Ref.~\cite{LerDecBer00} for
CTAB(0.3M)/${\rm NaNO}_3\rm {(1.79M)}$/H$_2$O at micellar volume
fraction $\phi=11\%$. This effect is much more pronounced in other,
more concentrated systems that are near an underlying (zero-shear)
isotropic-nematic (I-N) transition ($\phi\approx
30\%$)~\cite{schmitt94,BRL98,berret94a}.

In a cylindrical Couette geometry, this upward slope is qualitatively
consistent with the inhomogeneous stress arising from the cell
curvature: as the high-shear band at the inner cylinder expands
outward with increasing applied shear rate, the applied torque must
increase to ensure that the interface between the bands stays at the
selected stress $\sigmas$~\cite{olmsted99a}.  However a more generic
explanation, independent of geometry, is that the shear banding
transition is coupled to concentration~\cite{schmitt95,olmstedlu97}.
In this case, the properties of each phase must change as the applied
shear rate is tracked through the coexistence regime, because material
is redistributed between the bands as the high shear band grows to
fill the gap.

Generically, one expects flow to be coupled to concentration in
viscoelastic solutions where the different constituents (polymer and
solvent) have widely separated relaxation
timescales~\cite{brochdgen77,HelfFred89,DoiOnuk92,milner93,WPD91,beris94,SBJ97,tanak96}.
This was explained by Helfand and Fredrickson (HF)~\cite{HelfFred89}
as follows.  In a sheared solution, the parts of an extended polymer
molecule (micelle for our purposes) in regions of lower viscosity
will, upon relaxing to equilibrium, move more than the parts mired in
a region of high viscosity and concentration. A relaxing molecule
therefore on average moves towards the higher concentration region.
This provides a positive feedback mechanism whereby micelles can move
{\em up} their own concentration gradient, and leads to flow-enhanced
concentration fluctuations perpendicular to the shear compression
axis.
This was observed in steadily sheared polymer solutions in the early
1990's~\cite{WPD91}. In a remarkable paper, Schmitt
\etal~\cite{schmitt95} discussed the implications of this feedback
mechanism for the onset of flow instabilities. Strongly enhanced
concentration fluctuations were subsequently observed in the early
time kinetics of the shear banding instability in
Ref.~\cite{DecLerBer01}.

Recently, therefore, we introduced a model of concentration-coupled
shear banding~\cite{FieOlm02,FieOlm02b} by combining the diffusive
Johnson Segalman (d-JS) model~\cite{johnson77,olmsted99a} with a
two-fluid approach~\cite{brochdgen77,milner91,deGen76,Brochard83} to
concentration fluctuations. This ``d-JS-$\phi$" model does not address
the microscopics of any particular viscoelastic system, but instead
should be regarded as a minimal model that combines (i) a constitutive
curve like that of semidilute wormlike micelles (Fig.~\ref{fig:schem})
with (ii) the non-local (interfacial) terms required for selection of
a unique banded state~\cite{lu99} and (iii) a simple approach to
concentration coupling.

In Refs.~\cite{FieOlm02,FieOlm02b}, we examined the linear stability
of initially homogeneous shear states in this \model\ model with
respect to coupled fluctuations in shear rate $\gdot$, micellar strain
$\tens{W}$ and concentration $\phi$. We thereby calculated the
``spinodal'', inside which such homogeneous states are unstable.  We
also calculated the selected length scale at which inhomogeneity first
emerges during startup flows in the unstable region. In the limit of
zero concentration coupling, the unstable region coincides with that
of negative slope in the homogeneous constitutive curve, as expected;
but no length scale is selected during startup. Concentration coupling
enhances this instability at short length scales. It thereby broadens
the region of instability, and selects a length scale at which
inhomogeneity must emerge.

In the present paper, we compute the corresponding steady-state flow
phase diagram (the ``binodals'' and their tie-lines). As far as we are
aware, this is the first concrete calculation aimed at qualitatively
describing concentration-coupled shear banding for systems such as
semi-dilute wormlike micelles.  We start in Sec.~\ref{sec:background}
by describing the experimental background in more detail.  We also
compare our present calculation with the only other existing one for
concentration-coupled shear banded states, in concentrated solutions of
rigid rods~\cite{olmsted99c}. In Sec.~\ref{sec:model} we summarize our
\model\ model. We then review our results for the spinodal onset of
instability in Sec.~\ref{sec:spinodals}. In Sec.~\ref{sec:numerics} we
describe our numerical procedure for computing the banded steady
states, with brief discussion of our careful study of mesh and finite
size effects.  We then (Sec.~\ref{sec:results}) present our results
for the flow phase diagrams and shear-banded profiles. We conclude in
Sec.~\ref{sec:conclusion}.

\section{Experimental background; theoretical context}
\label{sec:background}

In this section, we discuss in more detail the experimental evidence
for concentration coupling in the shear banding of wormlike micelles.
We survey both (i) concentrated systems near the zero-shear I-N phase
transition and (ii) semidilute systems, in which underlying nematic
interactions are likely to be less important.  Correspondingly, we
compare the present calculation (aimed at the semi-dilute systems)
with an earlier calculation of flow phase diagrams in concentrated
solutions of rigid rods (near the I-N transition)~\cite{olmsted99c}.

The earliest observations of an upwardly sloping stress plateau in
wormlike micelles were made by Schmitt \etal\ \cite{schmitt94} and
Berret \etal~\cite{berret94a,BRL98}. Schmitt \etal\ \cite{schmitt94}
studied CpClO$_3$/NaClO$_3$(0.05M)/H$_2$O at the high micellar volume
fraction $\phi\approx 31\%$, just below the onset of the I-N
transition at $\phil=34\%$.  
In the steady-state flow curve, the stress increased smoothly up to
the critical shear rate $\gdotl$, where it showed a pronounced
downward kink before curving upward again for $\gdot>\gdotl$
(qualitatively like B'F' in Fig.~\ref{fig:schem}). SANS measurements
confirmed a superposition of nematic and isotropic contributions in
this regime $\gdot>\gdotl$, with the nematic contribution rising
linearly from zero at $\gdot=\gdotl$.

Berret \etal~\cite{berret94a,BRL98} studied CpCl/hexanol/NaCl for
several micellar volume fractions, again at a volume fraction just
below the onset of the zero-shear I-N transition ($\phil\approx
32\%$). The overall height of the coexistence plateau (which again
sloped upwards in $\gdot$) was found to fall with increasing
surfactant concentration $\phi\to\phil$, extrapolating to zero at
$\phi\gae\phil$, which is already biphasic in zero shear.  They also
found an increasing nematic contribution to SANS patterns for
increasing shear rates above $\gdotl$. They further used the SANS data
to show that the nematic (high shear) band was more concentrated than
the low shear band.

As noted above, the majority of existing calculations of shear-banded
states have assumed uniform concentration. An important exception is
the calculation of Olmsted and Lu~\cite{olmsted99c}. Although this
model was aimed at concentrated solutions of rigid rods, it broadly
captured some of the experimental phenomenology for the concentrated
($\phi\approx 30\%$) wormlike
micelles~\cite{schmitt94,berret94a,BRL98}.  For example, the overall
height of the coexistence plateau $\sigmas$ increased from zero as the
concentration was reduced below the threshold $\phil$ of the
zero-shear I-N biphasic regime.  The coexistence plateau sloped upward
markedly in shear rate. In further agreement with experiment, the high
shear (nematic) phase had a higher volume fraction of rods.  It should
be noted that model of Ref.~\cite{olmsted99c} was explicitly aimed at
concentrated systems, which in zero shear are already close to the I-N
transition: hence, the dynamics of the relevant order parameter
$\tens{Q}$ was driven by a free energy that already contained a phase
transition. In contrast, the simple free energy $F_{\rm e}(\tens{W})$
we consider below has no underlying phase transition, and flow-induced
efects are driven by convective, rather than dissipative
(relaxational) dynamics.

\begin{figure}[h]
 \includegraphics[scale=0.35]{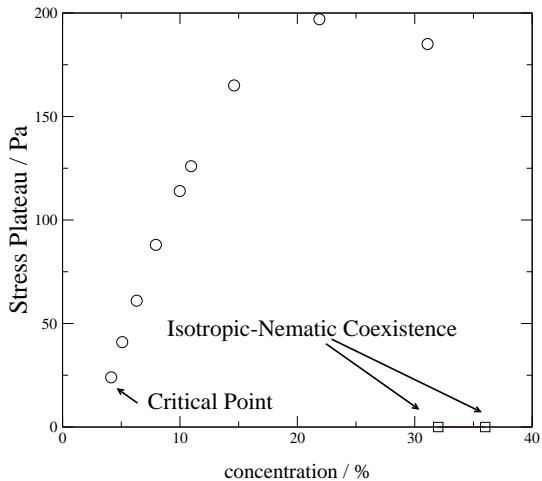}
\caption{Height of the coexistence plateau in the system
  CpCl/NaSal/brine. The 5 leftmost points are taken from the data of
  Ref.~\cite{berret94b}. The righthand point represents the zero-shear
  biphasic regime of this system, and is in accordance with the
  extrapolation of $G(\phi)/\Sigma_{\rm sel}$ in Ref.~\cite{berret94b}
  (see main text for details).
\label{fig:coexistence} } 
\end{figure}

There have also been several experimental studies of concentration
dependence in the shear banding of more dilute wormlike micellar
solutions. For example, Berret \etal~\cite{berret94b} investigated the
non-linear rheology of CpCl/NaSal/brine in the concentration range
$5\%-20\%$, well below the I-N$_{\rm c}$ transition at $\phi\approx
36\%$. (N$_{\rm c}$ is ``nematic calamitic''.) In contrast to the more
concentrated ``prenematic'' systems, the plateau height $\sigmas$
decreased with {\em decreasing} concentration: see the left 5 points
in Fig.~\ref{fig:coexistence}.  The width of the plateau also
decreased so that the difference $\gdot_{\rm h}-\gdot_\ell$ fell to
zero at a critical point $\phic,\gdotc,\Sigma_{\rm c}$ (leftmost point
in Fig.~\ref{fig:coexistence}). (These trends are the same as those in
Fig.~\ref{fig:alpha1.0e-3_gamma0.0:d} below.) In contrast the {\em
  scaled} plateau height $\sigmas/G(\phi)$ (where $G$ is the plateau
modulus) decreased with {\em increasing} concentration and
extrapolated to zero in the zero-shear biphasic (I-N) regime at
$\phi\gae\phil\approx 33\%$. According to this extrapolation (which is
actually well beyond the final data point at $\phi\approx 22\%$), and
in the absence of a divergence in $G(\phi)$, the unscaled plateau
height $\sigmas$ must itself fall to zero at $\phi\approx 33\%$
(rightmost data point in Fig.~\ref{fig:coexistence}), consistent with
the behaviour of the concentrated systems discussed above. To
summarise, the plateau height $\sigmas$ appears to be a non-monotonic
function of concentration, increasing with $\phi$ through the studied
regime $5\%<\phi<20\%$ before (probably) falling to zero in the
zero-shear biphasic $I-N$ regime $\phi\approx 33\%$.

Although this experiment showed that shear banding depends on the
{\em overall} concentration of the solution, there is relatively little
evidence for concentration-{\em coupling} (\ie\ concentration
differences between the bands) in such dilute systems, far from the
I-N transition.  Indeed, the experiment just described revealed no
discernible upward slope in the coexistence plateau. We are not aware
of any measurements of concentration differences between the
coexisting shear bands in such systems.  Nonetheless, recent
experiments on CTAB(0.3M)/${\rm NaNO}_3\rm {(1.79M)}$/H$_2$O at
$\phi=11\%$~\cite{LerDecBer00}
did reveal a stress plateau with slight upward slope.  Along with the
generic expectation that flow should be coupled to concentration in these
viscoelastic solutions, this suggests that an explicit calculation of
concentration difference in the shear bands of systems far from an I-N
transition might be worthwhile.

In this paper, therefore, we present the first such calculation, using
our \model\ model~\cite{FieOlm02,FieOlm02b}.  In contrast to the work
of Olmsted \etal\ \cite{olmsted99c} for concentrated rigid rods,
shear banding in the \model\ model is not due to any underlying nematic
feature of the elastic free energy $F^{\rm e}(\tens{W})$. Instead the
instability results mainly from the non-linear effects of shear (the
intrinsic constitutive curve has a region of negative slope), though
it can be strongly enhanced by concentration coupling in systems close
to an underlying Cahn-Hilliard (CH) demixing instability (governed by
the osmotic free energy $F^{\rm o}(\phi)$).  Indeed, the \model\ model
captures a broad crossover between (i) instabilities that are mainly
mechanical (governed by the negative slope of the flow curve) and (ii)
instabilities that are essentially CH demixing (governed by $F^{\rm
  o}(\phi)$), but now triggered by shear.  [Likewise, in practice
there should be no sharp distinction between concentrated micelles
with an underlying nematic feature in $F^{\rm e}(\tens{W})$ on the one
hand and ``non-nematic'' (more dilute) systems on the other: any more
refined model should allow a smooth crossover between the two cases.
This will be the focus of a future publication~\cite{FieOlm02f}.]

\section{Model}
\label{sec:model}

In this section we outline the \model\ model, which couples shear
banding instabilities to concentration in a simple way by combining
the non-local Johnson-Segalman (d-JS) model~\cite{johnson77} with a
2-fluid framework~\cite{brochdgen77,milner93} for concentration
fluctuations. While this description is self-contained, readers
are referred to Ref.~\cite{FieOlm02b} for fuller details.

\subsection{Free energy}

In a sheared fluid, one cannot strictly define a free energy.
Nonetheless, for realistic shear rates, many internal degrees of
freedom of a polymeric solution relax quickly on the timescale of the
moving constraints and are therefore essentially equilibrated.
Integrating over these fast variables, one obtains a free energy for a
given fixed configuration of the slow variables. For our purposes, the
relevant slow variables are the fluid momentum and micellar
concentration $\phi$ (which are both conserved and therefore truly
slow in the hydrodynamic sense), and the micellar strain $\tens{W}$
that would have to be reversed in order to relax the micellar stress
(which is slow for all practical purposes):
\begin{equation}
{W}_{\alpha\beta}=\frac{\partial{R}'_{\alpha}}{\partial
  \vect{R}}\cdot\frac{\partial {R}'_{\beta}}{\partial
  \vect{R}}-\delta_{\alpha\beta}
\end{equation}
where $\delta\vect{R}'$ is the deformed vector corresponding to the
undeformed vector $\delta\vect{R}$. 

The resulting free energy is assumed to comprise separate 
osmotic and elastic components,
\begin{equation}
\label{eqn:free_energy_total}
F=F^{\rm o}(\phi)+F^{\rm e}(\tens{W},\phi).
\end{equation}
%
%
The osmotic component is
\bea 
\label{eqn:free_energy}
F^{\rm o}(\phi)&=&\int d^3x \left[f(\phi)+\frac{g}{2}(\nablu \phi)^2\right]\nonumber\\
&\approx& \tfrac{1}{2} \int d^3q\, (1+\xi^2 q^2)f''|\phi(q)|^2,
\end{eqnarray}                             %
where $f''$ is the osmotic susceptibility and $\xi$ is the
equilibrium correlation length for concentration fluctuations.
The elastic component is
\begin{equation}
\label{eqn:elastic_fe}
F^{\rm e}(\tens{W},\phi)=\tfrac{1}{2}\int d^3x\,G(\phi) {\rm tr} \left[\tens{W}-\log(\tens{\delta}+\tens{W})\right]
\end{equation}
in which $G(\phi)$ is the micellar stretching modulus. 

\subsection{Dynamics}

The basic assumption of the two-fluid model is a separate force
balance for the micelles (velocity $\vm$; volume fraction $\phi$) and
the solvent (velocity $\vs$) within any element of solution.

The micellar force balance equation is:
\bw
\begin{equation}
\label{eqn:micelle}
\rho_{\rm m}\,\phi\,\left(\partial_t+\vm.\nablu\right)\vm=\nablu.G(\phi)\,\tens{W}-\phi\,\nablu \frac{\delta F(\phi)}{\delta\phi}+2\,\nablu.\,\phi\, \etam\, \Dm^{0} -\zeta(\phi)\,\vrel-\phi\nablu p.
\end{equation}
In this equation, $G(\phi)\tens{W}\equiv
2G(\phi)\tens{W}.\tfrac{\delta F}{\delta\tens{W}}$ is the viscoelastic
micellar backbone stress due to deformation of the local molecular
strain, while the osmotic stress $\tfrac{\delta F^{\rm o}}{\delta
  \phi}$ results from direct monomeric interaction.  The Newtonian
stress $2\phi\, \etam\, \Dm^{0}$ describes fast micellar processes
(\eg\ Rouse modes) with $\Dm^{0}$ the traceless symmetric micellar
strain rate tensor.  The force $\zeta\vrel$, where $\vrel=\vm-\vs$,
impedes relative motion; $\zeta$ is the drag coefficient
(Eq.~\ref{eqn:relative}).  Incompressibility determines the pressure
$p$.

Likewise, the solvent force balance comprises the Newtonian viscous
stress, the drag force (equal and opposite to the drag on the
micelles), and the hydrostatic pressure:
\begin{equation}
\label{eqn:solvent}
\rho_{\rm s}(1-\phi)\left(\partial_t+\vs.\nablu\right)\vs=2\nablu.\,(1-\phi)\,
\etas\, \Ds^{0}+\zeta(\phi)\vrel-(1-\phi)\nablu p.
\end{equation}
%

Equations~\ref{eqn:micelle} and \ref{eqn:solvent} contain the basic
assumption of ``dynamical asymmetry'', \ie\ that the viscoelastic
stress acts only on the micelles and not on the solvent. Adding them,
and assuming equal mass densities $\rho_{\rm m}=\rho_{\rm
  s}\equiv\rho$, we obtain the overall force balance equation for the
centre of mass velocity, $\vect{v}=\phi\vm+(1-\phi)\vs$:
\begin{equation}
\rho \left(\partial_t+\vect{v}.\nablu\right)\vv\equiv D_t\vect{v}
=\nablu.G(\phi)\,\tens{W}-\phi\,\nablu\frac{\delta
  F(\phi)}{\delta\phi}+2\,\nablu.\,\phi\, \etam\,
\Dm^{0}+2\nablu.\,(1-\phi)\, \etas\, \Ds^{0} -\nablu p.    
\label{eqn:navier}
\end{equation}
Subtracting them (with each predivided by its own volume fraction), we
obtain an expression for the relative motion $\vrel=\vm-\vs$, which in
turn specifies the concentration fluctuations:
\begin{equation}
D_t\phi=-\nablu\cdot\phi(1-\phi)
\vrel=-\nablu\cdot\frac{\phi^2(1-\phi)^2}{\zeta(\phi)}\left[\frac{\nablu
    \cdot G(\phi)\tens{W}}{\phi}-\nablu\frac{\delta F}{\delta
    \phi}+\frac{2\,\nablu\cdot\,\phi\, \etam\,
    \Dm^{0}}{\phi}-\frac{2\nablu\cdot\,(1-\phi)\, \etas\,
    \Ds^{0}}{1-\phi}\right]
\label{eqn:relative}
\end{equation} 
which defines the micellar diffusion coefficient $D\equiv
f''(\phi)\phi^2(1-\phi)^2/\zeta$.  We have omitted negligible inertial
corrections to Eqs.~(\ref{eqn:navier}) and (\ref{eqn:relative})
\cite{FieOlm02b}.

The essence of the 2-fluid model is that the physically distinct
elastic and osmotic stresses appear together in the force-balance
equation~(\ref{eqn:navier}) and also in the generalised CH
equation~(\ref{eqn:relative}). This allows micellar diffusion in
response to gradients in concentration {\em and} in the viscoelastic
stress. We will see below that this gives rise to a positive HF
feedback between concentration and flow~\cite{HelfFred89}, allowing
micelles to diffuse up their own concentration gradient.

For the dynamics of the viscoelastic micellar backbone strain we use
the phenomenological d-JS model~\cite{johnson77,olmsted99a}: 
\begin{equation}
\label{eqn:JSd}
(\partial_t+\vm.\nablu)\tens{W}=a(\Dm.\tens{W}+\tens{W}.\Dm)+(\tens{W}.\Omm-\Omm.\tens{W})+2\Dm-\frac{\tens{W}}{\tau(\phi)}+\frac{l^2}{\tau(\phi)} \nablu^2 \tens{W},
\end{equation}
\ew
where $2\Omm=\nablu \vm - (\nablu \vm)^T$ with $(\nablu
\vm)_{\alpha\beta}\equiv \partial_{\alpha}(v_{\rm m})_\beta$.
$\tau(\phi)$ is the Maxwell time and $l$ is a length scale discussed
in Sec.~\ref{sec:interfaces} below. The slip parameter $a$
measures the non-affinity of the molecular deformation, \ie\ the
fractional stretch of the polymeric material with respect to that of
the flow field. For $|a|<1$ (slip) the intrinsic constitutive curve in
planar shear is capable of the non-monotonicity of
Fig.~\ref{fig:schem}.

We use Eqns.~\ref{eqn:navier},~\ref{eqn:relative} and~\ref{eqn:JSd},
together with the incompressibility condition, $\nablu.\vect{v}=0$, as
our model for the remainder of the paper.

\subsection{Flow geometry. Boundary conditions}
\label{sec:geometry}

We consider idealised planar shear bounded by infinite plates at
$y=\{0,L\}$ with $(\vect{v},\nablu v, \nablu \wedge \vect{v})$ in the
$(\hat{\vect{x}},\hat{\vect{y}},\hat{\vect{z}})$ directions.  We allow
variations only in the flow-gradient direction, and therefore set all
other derivatives to zero: $\partial_x\ldots=0$, $\partial_z\ldots=0$.
In appendix~\ref{app:fulleqns} we give all the relevant components of
the model equations~\ref{eqn:navier},~\ref{eqn:relative}
and~\ref{eqn:JSd} in this coordinate system.

The boundary conditions at the plates are as follows.  For the
velocity we assume there is no slip.  For the concentration we assume
\begin{equation}
\label{eqn:BCconc}
\partial_y \phi=\partial^3_y \phi=0,
\end{equation}
which ensures (in zero shear at least) zero flux of concentration at
the boundaries.  
Following Ref.~\cite{olmsted99a}, for the micellar strain we assume  
\begin{equation}
\label{eqn:BCstrain}
\partial_y W_{\alpha\beta}=0\;\forall\; \alpha,\beta.
\end{equation}
Conditions~\ref{eqn:BCconc} and~\ref{eqn:BCstrain} together ensure
zero concentration flux at the boundary even in shear.  For the
controlled shear rate conditions assumed throughout,
\begin{equation}
\label{eqn:constant_strain_rate}
\bar{\gdot}=\int_0^L dy \gdot(y)={\rm constant.}
\end{equation}

\subsection{The interfacial terms}
\label{sec:interfaces}

The model contains two different interfacial terms. The first is the
gradient term on the RHS of Eqn.~(\ref{eqn:JSd}).  The length $l$ in
this term could, for example, be set by the mesh size or by the
equilibrium correlation length for concentration fluctuations. Here we
assume the former, since the dynamics of the micellar conformation are
more likely to depend on gradients in molecular conformation than in
concentration.  Physically, one can interpret the gradient term in
equation~\ref{eqn:JSd} as resulting dynamically, from the diffusion of
stretched molecules across the interface~\cite{elkareh89}, or
statically, from nematic interactions between the micelles, or both.
There is, at present, no accepted theory for these gradient terms in
semi-dilute solutions.  The equilibrium correlation length $\xi$ of
course still enters our analysis through our second interfacial term,
in the osmotic free energy of Eqn.~\ref{eqn:free_energy}.

Together, $l$ and $\xi$ set the length scale of any interfaces.
Throughout this paper, we study the physical limit in which $l$ and
$\xi$ are small compared to the system size so that we have a sharp
interface connecting two bulk homogeneous phases.  In this case, the
solution to Eqns.~\ref{eqn:navier},~\ref{eqn:relative}
and~\ref{eqn:JSd} naturally fits the zero-gradient boundary
conditions, and is invariant under $y\to y/2$, $l\to l/2$ and $\xi\to
\xi/2$. Therefore, a simultaneous reduction in $l$ and $\xi$ by the
same factor only changes the overall length of the interface, and not
the values of the order parameters in each phase (which determines
the phase diagram). However the phase diagram does depend slightly on
the ratio $r=l/\xi$: below we will give results for $r=0,r=\infty$ and
$r=O(1)$.  This provides a concrete example of the early insight of Lu
and co-workers~\cite{lu99}, that the banded state must depend on the
nature of the interfacial terms.  This contrasts notably with
equilibrium phase coexistence, in which the dynamical equations are
integrable and therefore insensitive to interfaces.

\subsection{Model parameters}
\label{sec:parameters}

\begin{table}
  \begin{center}
    \begin{tabular}{|p{2.85cm}|c|c|c|}
     \hline
     {\small\bf Parameter} & {\small \bf Symbol} $Q$ & {\small \bf Value at $\phi=0.11$} & $\frac{d\log Q}{d\log\phi}$ \\
     \hline\hline
     {\small Rheometer gap} & $L$ & $0.15\mbox{\,mm}$  & 0 \\
     \hline
     {\small Maxwell time} & $\tau$ & $0.17\mbox{\,s}$ & 1.1 \\
     \hline
     {\small Plateau modulus} & $G$ & $232\mbox{\,Pa}$ & 2.2 \\
     \hline
     {\small Density} & $\rho$ & $10^3\mbox{\,kg\,m}^{-3}$ & 0 \\
     \hline
     {\small Solvent viscosity} & $\etas$ & $10^{-3}\mbox{\,kg\,m}^{-1}\mbox{s}^{-1}$ & 0 \\
     \hline
     {\small Rouse viscosity} & $\etam$ & $0.4\mbox{\,kg\,m}^{-1}\mbox{s}^{-1}$ & 0 \\
     \hline
     {\small Mesh size} & $l$ & 2.6$\times 10^{-8}\mbox{m}$ & -0.73 \\
     \hline
     {\small Diffusion coefficient} & $D$ & $3.5 \times 10^{-11}\mbox{m}^2\mbox{s}^{-1}$ &0.77 \\
     \hline
     {\small Drag coefficient} & $\zeta$ & $2.4\times10^{12}\mbox{kg\,m}^{-3}\mbox{\,s}^{-1}$ & 1.54 \\
     \hline
     {\small Correlation length} & $\xi$ & $6.0\times10^{-7}$m & -0.77 \\
     \hline
     {\small Slip parameter} & $a$ & $0.92$ & 0 \\
     \hline
    \end{tabular}
  \end{center}
 \caption{Experimental values of the model's parameters at volume
   fraction $\phi=0.11$ (column 3). Scaling laws for the dependence of
   each parameter upon $\phi$ (column 4). In most calculations we use
   the reference values of column 3 at $\phi=0.11$, then tune $\phi$
   using the scaling laws of column 4. Only where  stated do we allow
   the parameters to vary independently.  \label{table:parameters}} 
\end{table}

The \model\ model
(Eqns.~\ref{eqn:navier},~\ref{eqn:relative},~\ref{eqn:JSd}) has the
following parameters: the solvent viscosity $\etas$ and density
$\rho$; the plateau modulus $G$; the Maxwell time $\tau$; the Rouse
viscosity $\etam$; the mesh size $l$; the osmotic modulus $f''(\phi)$
and the equilibrium correlation length $\xi$ (recall
Eqn.~\ref{eqn:free_energy}); the drag coefficient $\zeta$ and the slip
parameter $a$.  We also need to know the typical rheometer gap, $L$.
A reference set of parameter values at $\phi=0.11$ is summarised in
table~\ref{table:parameters}.  These values were taken from experiment
or calculated using scaling arguments: see Ref.~\cite{FieOlm02b} for
details.  Note that explicit data is not available for $f''(\phi)$;
however dynamic light scattering gives the diffusion coefficient
\begin{equation}
  \label{eq:1}
  D\equiv \frac{f''(\phi)\phi^2(1-\phi)^2}{\zeta(\phi)}.
\end{equation}
In this paper we will be guided by these parameter values, but subject
to the following considerations.

First, we are only interested in steady states so for convenience can
take the limit of zero Reynolds number ($\rho=0$) and rescale the
kinetic coefficient $1/\zeta$ so that the diffusion time $L^2/D$ is of
order the Maxwell time.  These choices have no effect on the steady
state, but make our numerical calculation of it much more efficient
(by evolving the {\em dynamical}
equations~\ref{eqn:navier},~\ref{eqn:relative} and~\ref{eqn:JSd}).
Second, realistic interfaces are much narrower than the typical
rheometer gap, with $l$ and $\xi$ both of $O(10^{-4}L)$.  To resolve
such interfaces (allowing a minimal 10 numerical mesh points per
interface) would therefore require $O(10^5)$ grid points, while in
practice we are limited to $O(10^2)$.  We will therefore use
artificially large values of $l$ and $\xi$.  However this does not
affect the phase diagram, provided the interface is still small
compared with the gap size: see Sec.~\ref{sec:numerics} for details,
and Fig. \ref{fig:finiteSizeBeta} for a typical banded profile.
Finally, we artificially increase the Rouse viscosity $\etam$ by a
factor $50$ to ensure, again for numerical convenience, that the shear
rate of the high shear phase is not too large. This does
quantitatively change the phase diagram, but we checked that the
qualitative trends are not affected.

Exploring this large parameter space is a daunting prospect so we
shall not, in general, vary the parameters independently of each
other. Instead we simply tune the concentration $\phi$, relying on
known semi-dilute scaling laws for the $\phi$-dependence of the other
parameters (column 4 of table~\ref{table:parameters}).  However we
will, in separate $\phi-$sweeps, vary the degree of concentration
coupling, which is dictated by ratio of the elastic term
$\nablu.G(\phi)\tens{W}$ to the osmotic term $\nablu\frac{\delta
  F}{\delta \phi}$ and which we encode in the parameter
\begin{equation}
  \label{eq:2}
  \alpha\equiv \frac{G'(\phi=0.11)}{2f''(\phi=0.11)}
\end{equation}
(where a prime denotes a derivative).  In other $\phi-$sweeps we will
vary the characteristic interface widths $l(\phi=0.11)$ and
$\xi(\phi=0.11)$, to investigate any dependence of the phase diagram
on the ratio $l/\xi$ in the double limit $l/L\to 0$, $\xi/L\to 0$.  In
what follows, we adopt the convenient shorthand of $l$ for
$l(\phi=0.11)$ with the understanding that $l$ does actually varies
with $\phi$ according to the scaling given in
table~\ref{table:parameters}. We do likewise for $\xi$.

Throughout we rescale stress, time and length so that
$G(\phi=0.11)=1$, $\tau(\phi=0.11)=1$, and $L=1$.

\section{Intrinsic flow curves; spinodals}
\label{sec:spinodals}

\begin{figure}[tb]
  \includegraphics[scale=0.3]{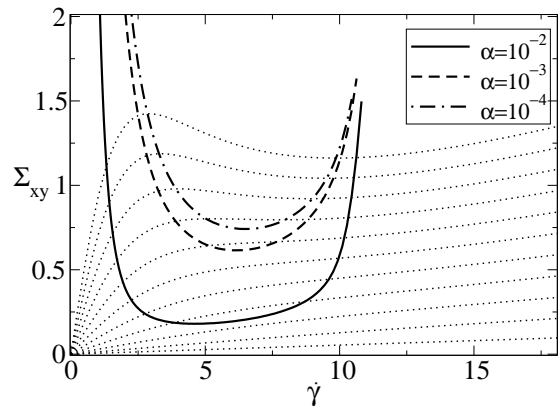}
\caption{Intrinsic flow curves (dotted lines) for
  $\phi=0.11,0.10\ldots0.01$ (downwards). Spinodals for concentration
  couplings $\alpha=10^{-2}, 10^{-3}, 10^{-4}$. 
\label{fig:spinodals}}
\end{figure}

The homogeneous intrinsic steady state flow curves
$\Sigma(\gdotb,\phib)=G(\phib)W_{xy}+\eta(\phib)\gdotb$ that satisfy
$\partial_t\vv=\partial_t\phi=\partial_t\tens{W}=0$ are shown as
dotted lines in Fig.~\ref{fig:spinodals}. (The average viscosity
$\eta(\phi)\equiv\phi\etam+(1-\phi)\etas$.)  The region of negative
slope ends at a ``critical'' point $\phib_{\rm c}\approx 0.015$.
CPCl/NaSal in brine~\cite{berret94b} shows the same trend. For
completeness, in App.~\ref{app:homogen} we give analytical results for
the steady state conditions in homogeneous shear flow.

In Ref.~\cite{FieOlm02,FieOlm02b} we linearised in fluctuations about
these homogeneous states to find the spinodal region in which the
homogeneous states are unstable.  The spinodals are shown in
Fig.~\ref{fig:spinodals} for different levels concentration coupling,
$\alpha$.  

In the limit of zero concentration coupling $\alpha\to 0$,
fluctuations in the ``mechanical variables'', $\tens{W}$ and $\gdot$
decouple from those in concentration, and are unstable in the region
of negative constitutive slope, as expected. Separately, the
concentration could have its own Cahn-Hilliard demixing instability,
when the diffusion coefficient $D<0$; however we are interested only
in flow-induced instabilities and set $D>0$ throughout.  For finite
$\alpha>0$, the region of mechanical instability is broadened by
coupling to the concentration fluctuations, as seen in
Fig.~\ref{fig:spinodals}. This can be understood as follows.  Consider
the first term in the square brackets of Eqn.(\ref{eqn:relative}).
This causes micelles to move up gradients in the viscoelastic stress
$\tens{W}$, thereby increasing the concentration in stressed regions.
If $G'(\phi)>0$ (assumed here), the increased concentration causes the
stress to increase further, closing a positive HF~\cite{HelfFred89}
feedback loop whereby the micelles can diffuse {\em up} their own
concentration gradient.

\section{Numerical details}
\label{sec:numerics}

In this section, we outline our numerical procedure for solving the
dynamical equations~\ref{eqn:navier},~\ref{eqn:relative}
and~\ref{eqn:JSd} and discuss our careful study of time-step, mesh
size and finite size effects. Readers who are not interested in these
issues can skip this section.

\begin{figure}[htb]
  \includegraphics[scale=0.3]{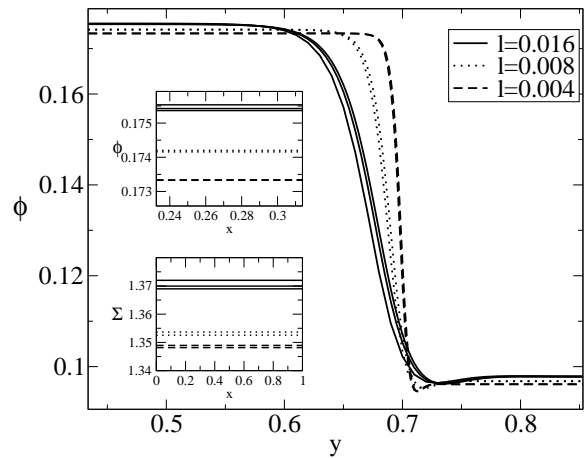}
  \caption{Main figure: steady banded concentration profile  at
    $\alpha=10^{-2}$, $\xi=0$, $\gdotb=7.0$, $\phib=0.15$. Solid lines
    $l=0.016$, for $(N_y,\dt)=(100,0.05),(200,0.0125),(400,0.0125)$,
    dotted lines $l=0.008$ for $(N_y,\dt)=(200,0.05),(400,0.003125)$,
    dashed lines $l=0.004$ for $(N_y,\dt)=(400,0.05),(800,0.05)$.
    Upper inset, the same data, enlarged in the left hand phase
    (decreasing $\phi$ with increasing $N_y$). Lower inset:
    corresponding selected stresses, for the same parameter values and
    mesh sizes (decreasing stress with increasing $N_y$).
\label{fig:finiteSizeBeta} } 
\end{figure}

We consider variations only in the flow gradient direction, in which we
discretise $y\in {0,1}$ on an algebraic grid $y_n=n/N_y$ for
$n=0,1\ldots N_y$.  We stored $\phi$ and $\tens{W}$ on these grid points.
The velocities $\vm$ and $\vs$ were stored on half grid points
$y_{n+1/2}$, and we used linear interpolation between the half and
full grid points.  Likewise we discretized time such that $t_n=n\Delta
t$.  We evolved the discretized
equations~\ref{eqn:navier},~\ref{eqn:relative} and~\ref{eqn:JSd} using
the Crank-Nicholson algorithm which is semi-implicit in time, with
centred space derivatives

\begin{figure*}
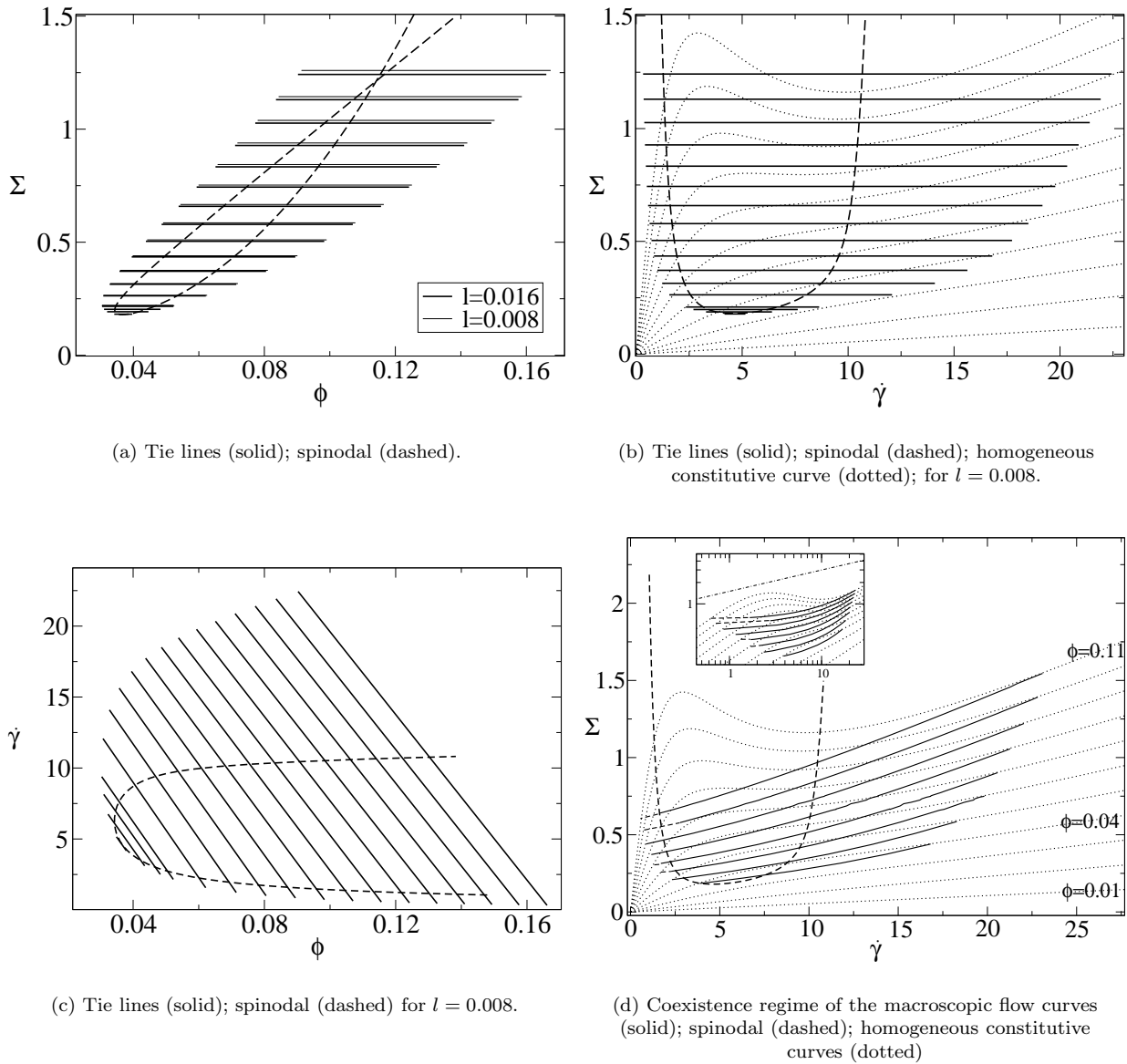

  \centering \subfigure[Tie lines (solid); spinodal (dashed).]{
    \includegraphics[scale=0.33]{sigmaVsPhi_alpha1p0e-3_gamma0p0.eps}
\label{fig:alpha1.0e-3_gamma0.0:a}
}
\subfigure[Tie lines (solid); spinodal (dashed); homogeneous
constitutive curve (dotted); for $l=0.008$. ]{ 
\includegraphics[scale=0.33]{sigmaVsgdot_alpha1p0e-3_gamma0p0.eps}
\label{fig:alpha1.0e-3_gamma0.0:b}
} \subfigure[Tie lines (solid); spinodal (dashed) for $l=0.008$.]{
\includegraphics[scale=0.33]{gdotVsPhi_alpha1p0e-3_gamma0p0.eps}
\label{fig:alpha1.0e-3_gamma0.0:c}
} \subfigure[Coexistence regime of the macroscopic flow curves (solid);
spinodal (dashed); homogeneous constitutive curves (dotted)]{
  \includegraphics[scale=0.3]{fc_alpha1p0e-3_gamma0p0_new.eps}
\label{fig:alpha1.0e-3_gamma0.0:d}
}
\caption{Phase diagrams and flow curves for $\alpha=10^{-2}$,
  $\xi=0.0$ for small $l/L$. (Recall that $l$ is actually a function
  of $\phi$: we are using the convenient shorthand of $l$ for the
  value $l(\phi=0.11)$.) (a) Thin (upper) solid lines: tie lines for
  $l=0.016,\ny=100,\dt=0.05$. Thick (lower) solid lines: tie lines for
  $l=0.008,\ny=200,\dt=0.05$. As described in the main text, we
  actually rescaled $l$ in the successive runs of each $\phib-$sweep
  (\ie\ as $\phib$ was tracked from $0.15$ down to $\phic$) so that
  the interfacial width remained (approximately) constant throughout
  the sweep: the value of $l$ in the figure legends refers to the
  value used in the first run of the sweep, at $\phib=0.15$. (b,c)
  Solid lines: tie lines repeated in the $(\Sigma,\gdot)$,
  $(\Sigma,\phi)$ representations for $l=0.008,\ny=200,\dt=0.05$. (d)
  Solid lines: macroscopic flow curves for
  $\phib=0.11,0.10,\ldots0.04$ (downward). These flow curves were
  recontructed from the tie lines of the phase diagrams (using the tie
  lines shown in this figure, and some additional ones). Because we
  have only calculated tie lines for discrete values of $\Sigma$, in
  some cases the reconstructed flow curves stop short of the
  single-phase region, and have been continued by eye with a dashed
  line. The inset in (d) shows the same data, but on a log-log plot.
  The experimentally observed slope $0.3$ is marked as a dot-dashed
  line for comparison.  The spinodal is shown in each of Figs a-d as a
  dashed line. In (b,d) the thin dotted lines are the intrinsic
  (homogeneous) constitutive curves for $\phib=0.11,0.1\ldots0.01$
  (downwards).}
\label{fig:alpha1.0e-3_gamma0.0}
\end{figure*}

For each run, we seeded an initial profile that was either homogeneous
up to a small random contribution, or inhomogeneous according to
$\phi=\bar{\phi}[1+\Delta\cos(\pi x)]$ (with $\Delta\approx 0.1$).  We
then evolved the discretized
equations~\ref{eqn:navier},~\ref{eqn:relative} and~\ref{eqn:JSd} under
an imposed wall velocity until a steady banded state was reached. We
checked that the homogeneous phases between the interfaces were
insensitive to the initial conditions.  However, for the random
initial condition several bands could form (and did not coarsen over
any accessible timescale). Therefore in most runs we used the
co-sinusoidal initial profile, to conveniently obtain just two bands
(as in Fig.~\ref{fig:profiles_compared}).

For the dynamics to be independent of time-step, a very small
time-step has to be used. However the numerically-attained steady is
much less sensitive hence allowing much larger time-steps.  A typical
steady state changes by less than $10^{-3}\,\%$ for a factor-two
reduction in timestep \ref{fig:finiteSizeBeta}. For the special case
of $\xi=0$, timesteps $\Delta t \propto \ny^{-2}$ can be used, since
the highest spatial derivative is second order. For $\xi\neq 0$, we
have a fourth order derivative in Eqn.~\ref{eqn:relative} and much
smaller timesteps $\Delta t\propto \ny^{-4}$ must be used.

In all our calculations, we are interested in the physical limit where
the interface width is much smaller than the rheometer gap.  This
creates a delicate balance, since narrow interfaces require a very
fine grid. Therefore we adopted the following procedure.  For any
fixed value of the interfacial lengthscales $l$ and $\xi$, we
performed several runs with progressively finer meshes (but always
with a small enough time-step) until the shear banded profile and
selected stress didn't depend on the mesh. This is quite easy to
achieve: a typical steady state presented below changes by less than
$0.1\,\%$ upon doubling the number of mesh points. We then reduced $l$
and $\xi$ (in fixed ratio) until the order parameters in the
homogeneous phases changed by less than $0.5\%$ upon further halving
of $l$ and $\xi$ (but always ensuring convergence with respect to the
number of grid points).  A sample study of these issues is presented
in Fig.~\ref{fig:finiteSizeBeta} for the special case $\xi=0$.

\section{Results}
\label{sec:results}

We now present our results for the steady-state flow phase diagrams,
flow curves and shear banded profiles. Because one of our aims is to
show that the shear-banded state depends on the nature of the
interfacial terms, we consider three separate cases: (A) interfacial
terms only in the viscoelastic constitutive equation~\ref{eqn:JSd}
($l\neq 0,\,\xi=0$, $r\equiv l/\xi=\infty$); (B) interfacial terms in
both the constitutive and concentration equations,~\ref{eqn:JSd}
and~\ref{eqn:relative} ($l\neq 0,\xi \neq 0$, $r=O(1)$); and (C)
interfacial terms only in the concentration
equation~\ref{eqn:relative} ($l=0,\xi\neq 0,r=0$).

\subsection{Interfacial terms only in the viscoelastic constitutive
  equation: $l\neq 0$, $\xi=0$.} 
\label{sec:onlyvisco}

In this section, we set the correlation length for concentration
fluctuations, $\xi$, to zero and consider small but non-zero values
the interfacial lengthscale $l$ in the constitutive
equation~\ref{eqn:JSd}.

\subsubsection{Flow phase diagrams}

\begin{figure}[htb]
\includegraphics[scale=0.3]{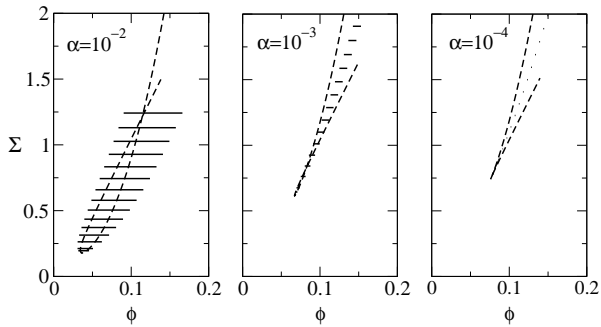}
\caption{Phase diagrams for three different degrees of coupling to
  concentration for $\xi=0$ and small $l/L$.
\label{fig:sigmaVsPhi_alpha1.0e-345_gamma0.0} } 
\end{figure}

For any given shear banded profile, the values of the order parameters
in each of the two homogeneous phases specify the two ends of one tie line
in the phase diagram. Analogously to equilibrium tie lines, the
concentrations and strain rates of the coexisting states are related
to the mean strain rate $\bar{\dot{\gamma}}$ and mean concentration
$\bar{\phi}$ by the lever rule,
\begin{align}
  \bar{\phi} &= \beta\phi_1 + (1-\beta)\phi_2 \\
  \bar{\dot{\gamma}} &= \beta\dot{\gamma}_1 + (1-\beta)\dot{\gamma}_2,
\end{align}
where $\beta$ is the volume fraction of material in state $(\phi_1,
\dot{\gamma}_1)$.  For each of several values of the concentration
coupling, $\alpha$, we calculated the full phase diagram via a
succession of shear startup runs, all at the critical shear rate
$\gdotc(\alpha)$ (determined from Fig.~\ref{fig:spinodals}), for
average concentrations ranging from $\phib=0.15$ down to the critical
value $\phic(\alpha)$.  For concentrations below the critical point
the response of the system is smooth as a function of stress.  In our
model, this arises because decreasing concentration reduces the
viscosity of the low shear rate branch faster than it reduces the
viscosity of high shear rate branch. Hence the stress maximum
decreases with decreasing concentration, disappearing when the stress
maximum vanishes.  Alternatively, in a more dilute system the plateau
modulus and Maxwell time are both smaller, and one expects a smaller
stress and higher strain rate at the onset of instability.

The results for $\alpha=10^{-2}$, which gives rather strong
concentration coupling, are shown in
Fig.~\ref{fig:alpha1.0e-3_gamma0.0}a,b,c.  Because the width,
$\delta$, of the interface in the banded state is set by $l$, but with
a prefactor that diverges at the critical point, in each successive
run we rescaled $l$ so that $\delta$ remained (approximately) equal to
its value ($\ll L$) in the first run at $\phib=0.15$.  We return below
to study the divergence of $\delta/l$ at the critical point.

To illustrate the finite size considerations of
Sec.~\ref{sec:numerics} (above), in
Fig.~\ref{fig:alpha1.0e-3_gamma0.0:a} we show the tie lines obtained
for two different (starting) values of $l$. All the results are
converged with respect to mesh fineness and timestep (not explicitly
shown), but the tie lines differ slightly between the two values of
$l$. However all seem to be consistent with one given binodal line: we
do not have any explanation for this apparent consistency.

To investigate the effect of reducing the coupling to concentration,
we repeat the phase diagram for $\alpha=10^{-2}$ alongside that for
$\alpha=10^{-3}$ and $\alpha=10^{-4}$ in
Fig.~\ref{fig:sigmaVsPhi_alpha1.0e-345_gamma0.0}.  As expected, the
concentration difference between the bands tends to zero as 
$\alpha\rightarrow0$.

\subsubsection{Flow curves}

\begin{figure*}
\subfigure[]{
    \includegraphics[scale=0.19]{reconstruct1.eps}
\label{fig:reconstruct:a}
}
\subfigure[]{
\includegraphics[scale=0.8]{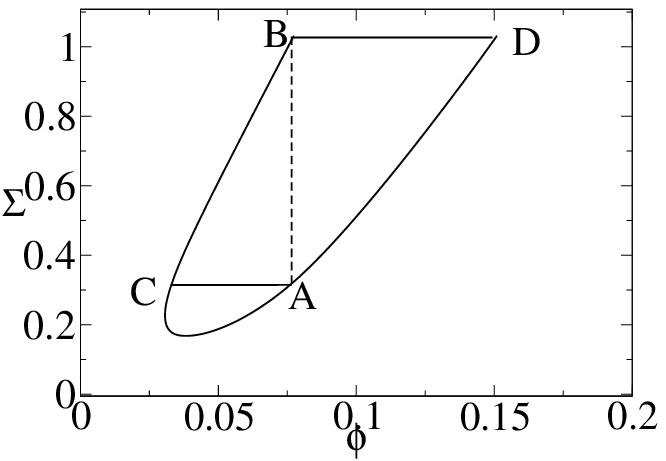}
\label{fig:reconstruct:b}
}
\subfigure[]{
\includegraphics[scale=0.8]{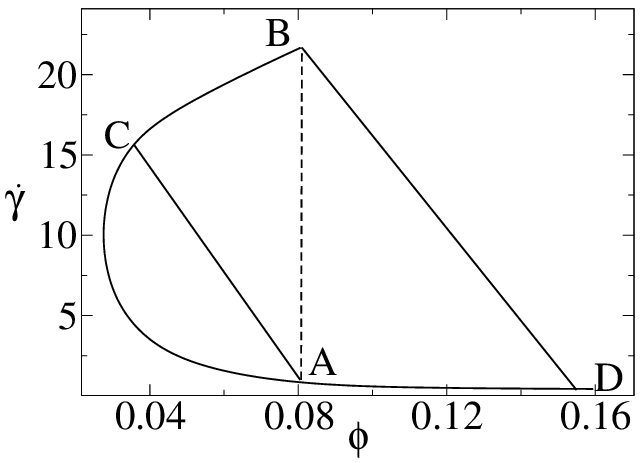}
\label{fig:reconstruct:c}
}
\caption{Reconstruction of the flow phase diagram from a family of
  macroscopic flow curves $\Sigma(\gdotb,\phib)$, measured for several
  different average concentrations $\phib$.  Consider the flow curves
  of Fig.~\ref{fig:reconstruct:a}. The curve that starts at A and ends
  at B is for an average concentration $\phib=0.08$. Points A and B
  are at the edge of the two-phase region.  Reading off the stress
  from Fig.~\ref{fig:reconstruct:a}, A and B give use two points on
  the binodal in Fig.~\ref{fig:reconstruct:b}.  Likewise reading off
  the strain rate, we get points A and B in
  Fig.~\ref{fig:reconstruct:c}. Repeating this for all the circles in
  Fig.~\ref{fig:reconstruct:a}, we can construct many points on the
  binodal in Figs.~\ref{fig:reconstruct:b} and
  \ref{fig:reconstruct:c}, which can then be interpolated over to give
  the full binodal. We now just need to specify the tie lines. In
  Fig.~\ref{fig:reconstruct:b} this is trivial: all tie lines are
  horizontal since the coexistence occurs at common stress (for
  gradient banding). In Fig.~\ref{fig:reconstruct:c}, to get the slope
  of the tie line that starts at $B$ we proceed by recalling that the
  tie line represents constant shear stress. Therefore we find another
  point, D, in Fig.~\ref{fig:reconstruct:a} that is at the same stress
  as point $A$, and read off its average strain-rate. Its average
  concentration is already known. This gives point $C$ in
  Fig.~\ref{fig:reconstruct:c}. Similarly, $D$ is the image of point
  $B$ at constant stress. Repeating this process we can fill in all
  the tie lines of the phase diagram.  }
\label{fig:reconstruct}
\end{figure*}

So far, we have discussed the flow phase diagrams.  Measurement of
these diagrams still presents an open challenge to experimentalists,
due to the difficulty in measuring the concentration of micelles in
each band (although SANS data has been used to estimate the bands'
concentrations in systems near the I-N transition~\cite{BRL98}). In
this section we discuss the macroscopic flow curves, which are
relativly easily measured using conventional bulk rheology.  However
it is important to realise that a set of flow curves
$\Sigma(\gdotb,\phib)$ measured for {\em several} values of $\phib$
actually contains the same information as the phase diagram:
reconstruction of the latter from the former is described in
Fig.~\ref{fig:reconstruct} A full set of flow curves could therefore
be used to check measurements of concentration differences.

\begin{figure}[htb]
\includegraphics[scale=0.3]{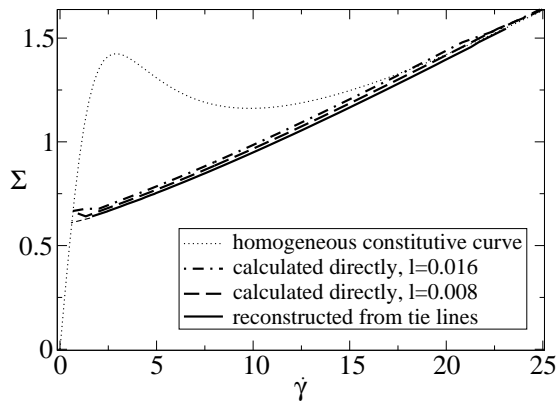}
\caption{Macroscopic flow curves for $\alpha=10^{-2}$ at
  $\phib=0.11$. Thick solid line:  reconstructed from the tie lines of
  the phase diagram. Dot dashed and dashed lines:  calculated by
  directly measuring the average stress and strain rate during a
  strain rate sweep for $l=0.016,\ny=100,\dt=0.05$ (dot-dashed) and
  $l=0.008,\ny=200,\dt=0.05$  (dashed). (The slight discrepancy
  between these three curves is discussed in the text.) The thin
  dotted line is the intrinsic (homogeneous) constitutive curve. 
\label{fig:calced_fc_alpha_1.0e-3_gamma0.0} } 
\end{figure}

\begin{figure}[htb]
\includegraphics[scale=0.3]{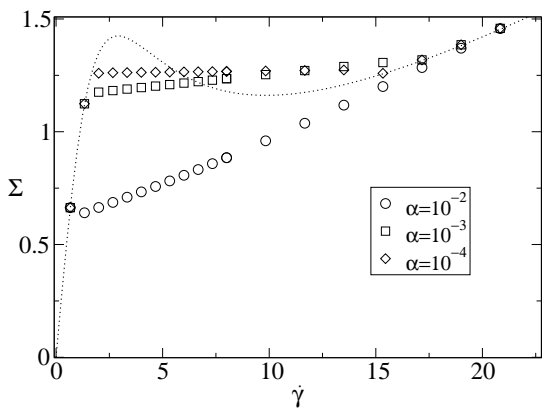}
\caption{Macroscopic flow curves (from direct measurements of the
  stress and strain rate) for three different degrees of coupling to
  concentration. 
\label{fig:fc_alpha1.0e-345_gamma0.0} } 
\end{figure}

\begin{figure}[htb]
\includegraphics[scale=0.3]{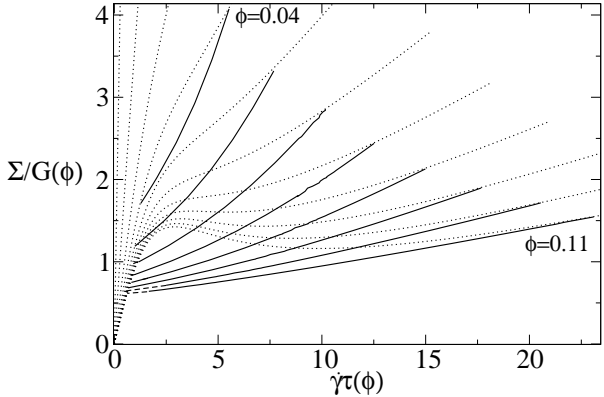}
\caption{Macroscopic flow curves as shown in
  Fig.~\ref{fig:alpha1.0e-3_gamma0.0:d} above, but now with the stress
  in units of $G(\phi)$ and the strain rate in units of $\tau(\phi)$.
\label{fig:fc_alpha1.0e-3_gamma0.0_scaled} } 
\end{figure}

In this work, we take the opposite approach for convenience, and
reconstruct the steady-state flow curves from the tie lines of the
phase diagram.  The results are shown in
Fig.~\ref{fig:alpha1.0e-3_gamma0.0:d}.  The inset shows the same data
on a log-log plot, to enable comparison with Ref.\cite{berret94a} in
which the coexistence plateau in a log-log representation was a
reasonably straight line (over the shear-rate range investigated) with
slope $0.3$.  Note that the results shown in
Fig.~\ref{fig:alpha1.0e-3_gamma0.0:d} are in units of $G(\phi=0.11)$
and $\tau(\phi=0.11)$. In Ref.~\cite{BRP94}, Berret replotted the flow
curves in units of $G(\phib)$ and $\tau(\phib)$, finding scaling
collapse of the family $\Sigma(\gdotb,\phib)/G(\phib)$ \versus\ 
$\gdotb\tau(\phib)$ in the low shear regime $\gdotb\to 0$. We do not
find this scaling collapse
(Fig.~\ref{fig:fc_alpha1.0e-3_gamma0.0_scaled}) because we have used
an artificially large high-shear Newtonian contribution $\eta\gdot$
for numerical convenience (recall Sec.~\ref{sec:parameters}): the
overall zero shear viscosity, $G(\phi)\tau(\phi)+\eta(\phi)$ therefore
does not scale as $G(\phi)\tau(\phi)$, even approximately.

To check the reconstruction of flow curves from the phase diagram, we
also explicitly calculated the flow curve at a single $\phib=0.11$. To
do this, we first performed a shear startup at a given $\gdotb$ in the
unstable region.  We then (without reinitialising the system)
decreased $\gdotb$ in steps to the edge of the coexistence regime,
ensuring that a steady state was reached before measuring the total
stress.  We then reinitialised the system and repeated the entire
procedure, but now with increasing $\gdotb-$jumps. The results are
shown in \ref{fig:calced_fc_alpha_1.0e-3_gamma0.0} for two different
values of $l$.  The slight discrepancy between the directly measured
flow curve ``plateaus'' (i.e. the inhomogeneouse part of the flow
curve) and those reconstructed from the tie lines is due to the finite
size of the interface $\delta$ relative to the cell $L$, and so is
smaller for the smaller value of $\delta/L$. The construction
described in Fig.~\ref{fig:reconstruct} implicitly assumes that
$\delta/L=0$.

As expected for this value of $\alpha$ (which gives a large
concentration difference between the bands;
Fig.~\ref{fig:alpha1.0e-3_gamma0.0:a}), the steady state flow curve
``plateau'' slopes strongly upwards in $\gdotb$.  In
Fig.~\ref{fig:fc_alpha1.0e-345_gamma0.0} we compare the (directly
measured) macroscopic flow curve for the three levels of concentration
coupling shown in Fig.~\ref{fig:sigmaVsPhi_alpha1.0e-345_gamma0.0}: as
expected, the slope of the flow curve tends to zero with the degree of
concentration coupling $\alpha$. 

The upturn in the measured flow curve at the edge of the coexistence
plateau (apparent at the lower binodal for $\alpha=10^{-2}$ in
Fig.~\ref{fig:fc_alpha1.0e-345_gamma0.0}) results again from the
finite value of $\delta/L$: the interface bumps into the edge of the
rheometer when one of the bands gets very narrow. We expect this
(steady-state) effect to be much less pronounced in experimental
systems, since realistic interfaces are much smaller than those used
in our numerical study. Only near a critical point, where the
interface becomes very broad (for fixed $l$), would we expect to see a
true steady-state bump at the edge of the plateau.  Nonetheless,
pronounced bumps are often apparent in data obtained via upward
strain-rate sweeps.  However in most cases this is likely to be a
metastable effect, so that the bump could be eliminated (or at least
reduced) by reducing the rate of the sweep~\cite{grand97}.

As noted in Sec.~\ref{sec:intro}, in a curved Couette geometry the
``plateau'' (B'F' of Fig.~\ref{fig:schem}) in the flow curve will
slope upwards due to the inhomogeneity of the stress field, even
without concentration coupling. It should be noted that all
calculations in this paper are for a planar shear geometry, and the
slope of our flow curves in the coexistence regime results solely from
concentration coupling. In fact, the slope in
Fig.~\ref{fig:alpha1.0e-3_gamma0.0:d} is far greater than one would
typically expect from curvature effects: for a Couette cell with
radius $R$ and gap $\delta R$, the stress measured at the inner
Couette wall would change by $\delta\Sigma/\Sigma=2\delta R/R$ over
the coexistence regime, and so too would the relative change in torque
through the coexistence ``plateau''.  The slope of
Fig.~\ref{fig:alpha1.0e-3_gamma0.0:d} would therefore require an
atypically large curvature of $\delta R /R\sim0.5$.

\subsubsection{Interfacial profiles; divergence of interface width at
  the critical point} 

\begin{figure}[htb]
\includegraphics[scale=0.4]{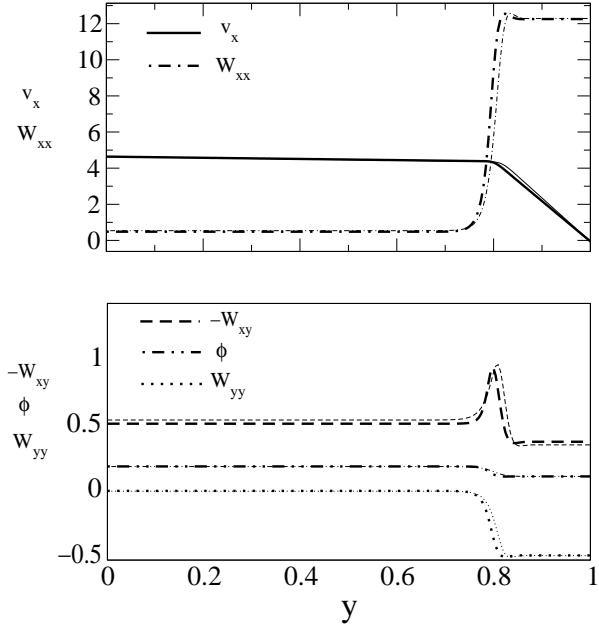}
\caption{Steady state shear banded profile at $\alpha=10^{-2}$,
  $\gdotb=4.64$, $\phib=0.15$ for two different ratios $r=l/\xi$. The
  thick lines are for $r=\infty$ ($l=0.008$, $\xi=0.0$), $\ny=200$,
  $\dt=0.05$, as considered in this section. The thin lines show the
  corresponding results for   $r=0.4$ ($l=0.008$, $\xi=0.002$),
  $\ny=200$, $\dt=0.00625$ (to be discussed in Sec.~\ref{sec:both}
  below), for comparison.  
\label{fig:profiles_compared} } 
\end{figure}

\begin{figure}[htb]
\includegraphics[scale=0.3]{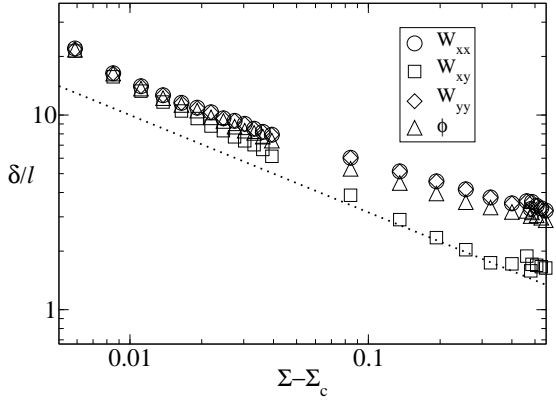}
\caption{Scaled interface width $\delta/l$ versus the distance from the critical stress $\Sigma-\Sigma_{\rm c}$. The dotted line is a power $-0.5$. 
\label{fig:widths_alpha1.0e-3_gamma0.0} } 
\end{figure}

We now turn to the interfacial profiles and widths.  A full steady
state banded profile for $\alpha=10^{-2}$ (corresponding to the
rightmost/uppermost tie line in
fig.~\ref{fig:alpha1.0e-3_gamma0.0}a,b,c) is shown by the thick lines
in Fig.~\ref{fig:profiles_compared}. As required, the interface is
smooth on the scale of the mesh, but sharp on the scale of the gap
size, \ie\ $L/\ny\ll \delta \ll L\equiv 1$ where $\delta$ is the width
of the interface. Note that the shear rate is negative across the gap
since we have chosen to move the wall at $y=0$; accordingly we have
plotted $-W_{xy}$, since $W_{xy}$ is antisymmetric in shear rate.
$-W_{wy}$ is rather small in the high shear band, as expected from the
underlying constitutive non-monotonicity. Meanwhile $W_{xx}$ is very
large, while $W_{yy}\approx -0.5$ (recall that $\tens{W}$ measures
deformation relative to the unit tensor $\tens{\delta}$): this
corresponds to the micelles being highly stretched along the flow
direction and is consistent with the experimental observation that the
first normal stress difference progressively increases throughout the
banding regime~\cite{rehage91}. The concentration is lower in the high
shear band, where $W_{yy}$ is smaller (more negative): this is a
direct result of the tendency of mielles to move up gradients in
$W_{yy}$, as determined by Eqn.~\ref{eqn:relative} above.

In fact the interface width, $\delta$, is slightly different for each
order parameter: we define it to be the distance between the two
points where the change in that order parameter between the two
homogeneous phases is $25\%$ and $75\%$ complete.  For a fixed value
of $l$ (which sets the overall scale of the interface width), $\delta$
diverges at the critical point (for each order parameter).  In
tracking $\phib$ down towards the critical point, therefore, we
continually rescaled $l$ to ensure that the interface width remained
approximately constant. In each case, we measured $\delta/l$, for each
of $W_{xy}$, $W_{xx}$, $W_{yy}$ and $\phi$: see
Fig.~\ref{fig:widths_alpha1.0e-3_gamma0.0}.  According to mean field
theory, the divergence should be of the form $\delta/l \sim
(\Sigma-\Sigma_{\rm c})^{-1/2}$. The power $-1/2$ is accordingly shown
in Fig.~\ref{fig:widths_alpha1.0e-3_gamma0.0} as a guide for the eye.

\subsection{Interfacial terms in both the viscoelastic constitutive
  equation, and in the concentration equation: $l\neq 0$, $\xi\neq
  0$.} 
\label{sec:both}


\begin{figure}[htb]
  \centering \subfigure[Tie lines]{
    \includegraphics[scale=0.32]{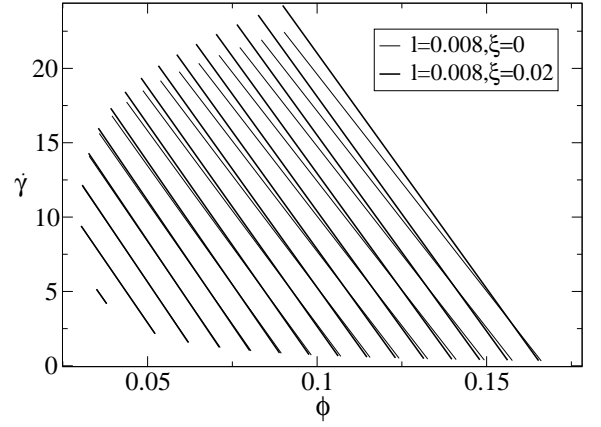}
    \label{fig:alpha1.0e-3_betaonly_and_both:a}
  } \subfigure[Partially reconstructed macroscopic flow curves
  (solid); homogeneous constitutive curve (dotted).]
  {\includegraphics[scale=0.32]{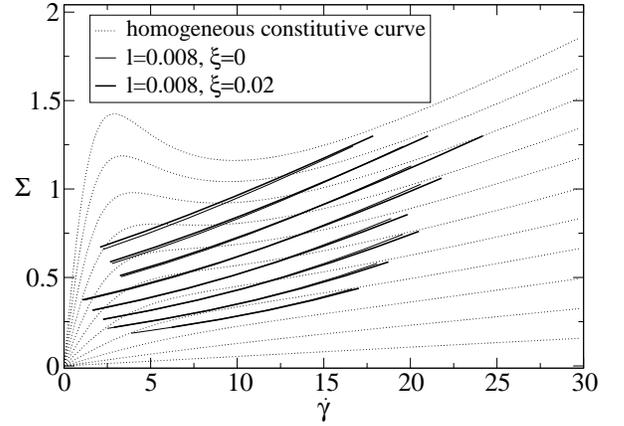}
    \label{fig:alpha1.0e-3_betaonly_and_both:b}
    }
  \caption{Phase diagrams and flow curves for $\alpha=10^{-2}$ and
    $r=0.4$ ($l=0.008$, $\xi=0.02$) with the corresponding data for
    $\alpha=10^{-2}$ and $r=\infty$ ($l=0.008$, $\xi=0$) for
    comparison. } 
  \label{fig:alpha1.0e-3_betaonly_and_both}
\end{figure}

We now study the effect of including interfacial gradient terms in the
concentration equation~\ref{eqn:relative} (so that now $\xi \neq 0$)
as well as in the viscoelastic equation~\ref{eqn:JSd}, $l\neq 0$.
Hence, while in the previous section we considered $r\equiv
l/\xi=\infty$, then, we now consider $r=O(1)$.  In
Fig.~\ref{fig:alpha1.0e-3_betaonly_and_both:a}, we give the phase
diagram for $r=0.4$. Comparing it with our results for $r=\infty$
(also shown in Fig.~\ref{fig:alpha1.0e-3_betaonly_and_both:a}), we see
that the slopes of the tie lines and the overall binodal both depend
quantitatively on $r$. [The difference between the results for
$r=\infty$ and $r=0.4$ is far greater than any ``error'' associated
with the fact that we are not quite in the limit $\dt\to 0$,
$l\ny\to\infty $, $\xi\ny\to\infty $, $l/L\to 0$ and $\xi/L \to 0$.]
This provides a concrete example of the fact that shear-banding
coexistence is determined by, and non-universal with respect to, the
interfacial terms~\cite{lu99}.  As noted above, this contrasts sharply
with the equilibrium case, in which the equations of motion are
integrable and so the phase diagram is independent of the interfacial
terms.  Although conceptually important, this dependence is in
practice rather weak: the overall features of the phase diagram are
unchanged. The critical point is unaffected.

In Fig.~\ref{fig:alpha1.0e-3_betaonly_and_both:b} we show the
corresponding macroscopic flow curves, reconstructed using the tie
lines of Fig.~\ref{fig:alpha1.0e-3_betaonly_and_both:a}. Because we
only calculated a few tie lines in this case, the recontruction is
rather sparse. Nonetheless, the slight difference between $r=\infty$
and $r=0.4$ is apparent. 

In Fig.~\ref{fig:profiles_compared}, we compare a full banded profiles
for $r=\infty$ and $r=0.4$. The slight dependence on $r$ is again
apparent.


\subsection{Interfacial gradient terms only in the concentration equation: $l=0$, $\xi\neq 0$.}
\label{sec:onlyconc}

\begin{figure*}
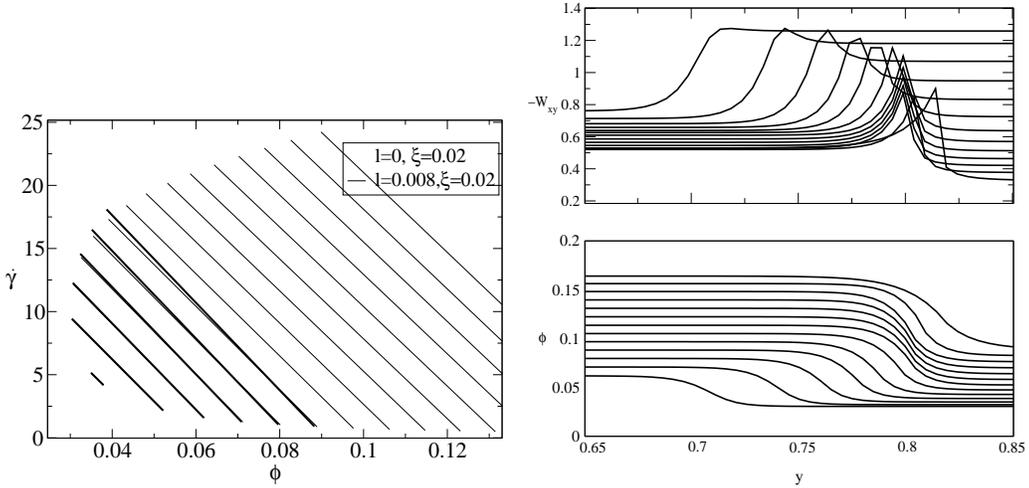
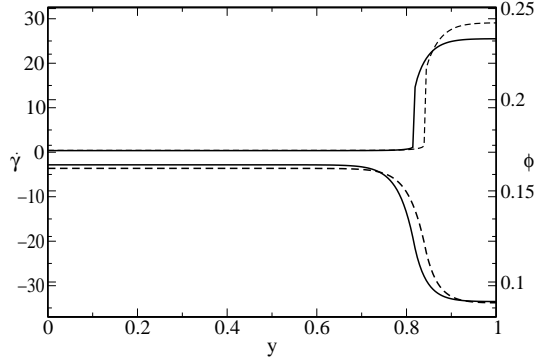

  \centering \subfigure[Tie lines, only shown for the case $l=r=0$
  when close to the critical point]{
    \includegraphics[scale=0.28]{gdotVsPhi_gammonly_and_both.eps}
\label{fig:gdotVsPhi_gammonly_and_both}
}
\subfigure[Steady-state profiles for initial condition
$\phi(y)=\phib+0.4\cos(\pi y)$: there is no selected smoothly banded
state ]{ \includegraphics[scale=0.33]{profiles_gammonly.eps}
\label{fig:profiles_gammonly}
}\\
\subfigure[Steady-state profiles for initial conditions
$\phi(y)=\phib+0.4\cos(\pi y)$ and $\phi(y)=\phib+0.7\cos(\pi y)$: the
steady state depends on the initial condition.]
{
\includegraphics[scale=0.25]{no_selection.eps}
\label{fig:no_selection}
}
\caption{(a) Phase diagram at $\alpha=10^{-2}$, for $l=0.0$,
  $\xi=0.02$, $r\equiv l/\xi=0.0$, shown with the corresponding data
  for $l=0.008$, $\xi=0.02$, $r=0.4$ for comparison. Tie lines are
  only shown near the critical point because for larger values of
  $\phib$, there is no uniquely selected, smoothly banded state. This
  is shown in Figs. b and c. In Fig.b the steady state profiles from
  left to right at fixed ordinate correspond to tie lines left to
  right in the upper Fig.(a). Fig (c) shows the steady state profile
  in $\gdot$ (upper two curves) and in $\phi$ (lower two curves) for
  $\phib=0.16$ and $\gdotb=4.66$ with initial condition
  $\phi(y)=\phib+0.4\cos(\pi y)$ (solid lines) and with
  $\phi(y)=\phib+0.7\cos(\pi y)$ (dashed lines): the ``selected''
  state depends upon the initial condition -- \ie\ there is no state
  selection for $l=0$ for stresses far enough above the critical
  point.}
\label{fig:alpha1.0e-3_gammonly}
\end{figure*}


Finally we set the interfacial length $l$ in the constitutive equation
equal to zero. The constitutive equation is now local, and the only
source of spatial gradients is the equilibrium correlation length for
concentration fluctuations (Eqns.~\ref{eqn:relative}
and~\ref{eqn:free_energy}): $r\equiv l/\xi=0$. In the absence of
concentration coupling, $\alpha=0$, it is known that there is no
uniquely selected, smoothly shear banded state when $l=0$~\cite{lu99}.
Here we investigate whether a smoothly banded state is selected for
$\alpha\neq 0$, by virtue of the interfacial terms in the
concentration equation.

Our numerics only gave a smoothly banded profile for stresses near the
critical point, even for the largest accessible values of $\xi$ and
$\ny$. The profiles shown from left to right in
Fig.~\ref{fig:profiles_gammonly} are progressively further above the
critical point. The tie lines correponding to the smooth profiles near
the critical point are shown in
Fig.~\ref{fig:gdotVsPhi_gammonly_and_both}, alongside the correponding
results at $r=0.4$ for comparison.  Consistent with the discussion of
non-universality in the previous section, the phase diagram for
$r=0.0$ is slightly different from that for $r=0.4$ (and is different
again from the case $r=\infty$; not shown).

For the spiky profiles, further from the critical point, the binodal
of the associated tie lines is irregular (not shown in
Fig.~\ref{fig:gdotVsPhi_gammonly_and_both}), suggesting that the
steady state is not uniquely selected.  In view of this, a natural
question is whether selection could occur in principle (but is
inaccessible with any realistic mesh due to the pronounced
non-monotonicity in $W_{xy}(y)$), or whether selection cannot occur,
even in principle. In Fig.~\ref{fig:no_selection} we show that the
steady state depends on the initial condition; so state selection
appears to be lost when $l=0$. This numerical observation is backed up
by the following analytical argument.

In steady state, the system must obey:
\begin{itemize}
\item The force-balance equation,
\begin{equation}
\label{eqn:simplenavier}
S(\gdot,\phi)\equiv G(\phi)W_{xy}[\gdot\tau(\phi)]+\etabar(\phi)\gdot=\Sigma={\rm const.}
\end{equation}
\item The (now local) constitutive equation, equation~(\ref{eqn:JSd}), 
\begin{equation}
\label{eqn:solveW}
W_{\alpha\beta}=W_{\alpha\beta}[\gdot \tau(\phi)] \;\;\mbox{for }\;\;\alpha\beta=xx,xy,yy.
\end{equation}
\item The steady-state of equation~\ref{eqn:relative}. For the
  purposes of this analytical argument we use a simplified version of
  this equation, which we believe still captures the essential physics:
\begin{equation}
\label{eqn:simpleconc}
0=\partial^2_y\left\{\Lambda-\partial_y^2\phi \right\}
\end{equation}
with
\begin{equation}
\label{eqn:defineF}
\Lambda=f'(\phi)-\frac{G(\phi)W_{yy}[\gdot\tau(\phi)]}{\phi}.
\end{equation}
%
Integrating Eqn.~\ref{eqn:simpleconc}
twice, and using the boundary conditions
$\partial_y\phi=0,\partial^3_y\phi=0$ for $y=0,L$, we obtain
\begin{equation}
\label{eqn:third}
\mu={\rm const.}=\Lambda-\partial_y^2\phi,
\end{equation}
where $\mu$ is an integration constant.
\end{itemize}
We now show that a solution satisfying Eqns.~\ref{eqn:simplenavier}
and~\ref{eqn:solveW} cannot in general simultaneously satisfy
Eqn.~\ref{eqn:third}.

\begin{figure*}[htb]
\includegraphics[scale=0.6]{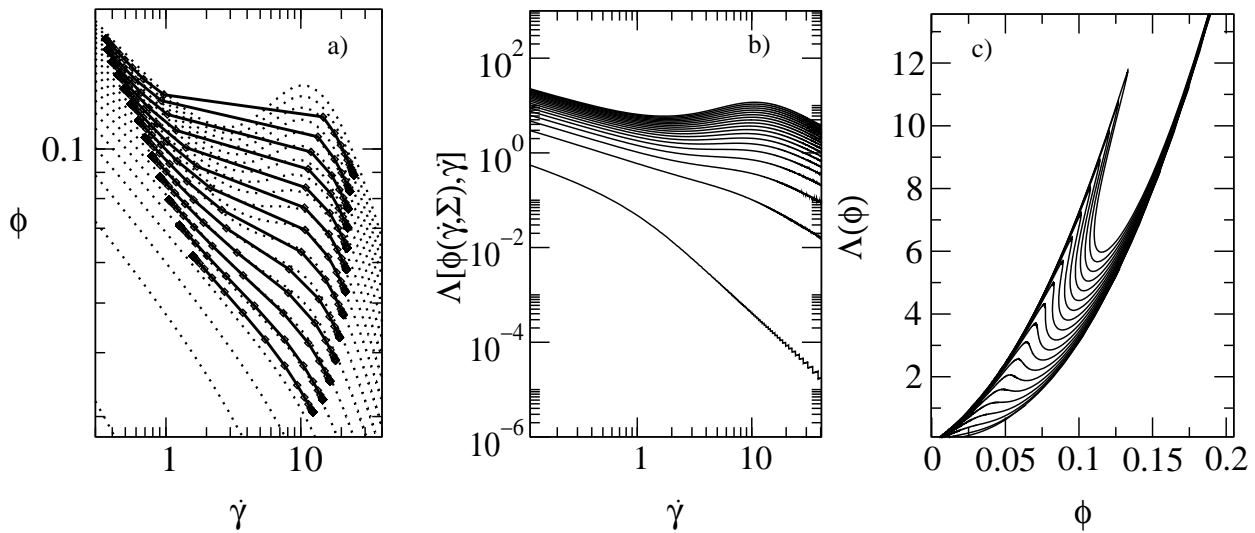}
\caption{a) Dotted lines: relation between $\phi$ and $\gdot$ for the
  case of a local constitutive equation, for several values of the
  shear stress, $\Sigma$, calculated using
  Eqns.~(\ref{eqn:simplenavier}) and~(\ref{eqn:solveW}). Solid lines:
  the results of our numerics, showing that the $\phi-$ profile cannot
  properly negotiate the interface, as described in the main text.
  (Note that the stresses used to generate the dotted and the solid
  lines differ slightly, but the overall trend is still clear.) b) The
  function $\Lambda$ of Eqn.~\ref{eqn:defineF} plotted \versus\ 
  $\gdot$ using the relation of Fig.a). c) $\Lambda$ replotted
  \versus\ $\phi$.
\label{fig:construction} } 
\end{figure*}
\begin{figure}[htb]
\includegraphics[scale=0.6]{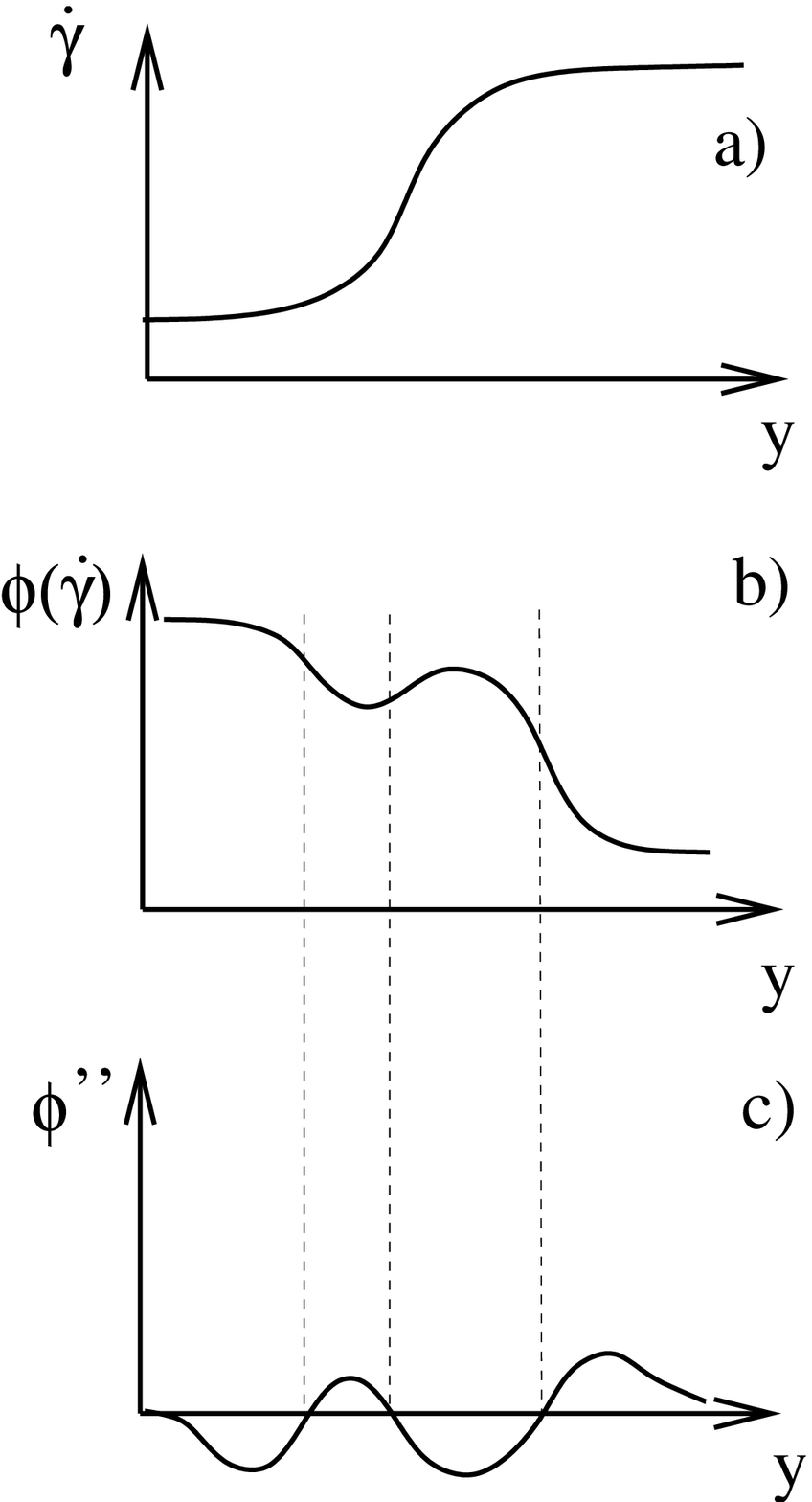}
\caption{Assuming that the shear rate varies monotonically across the
  interface (a), then for a relation $\phi(\gdot,\Sigma)$ of
  Fig.~\ref{fig:construction}a that is non-monotonic, the
  concentration $\phi$ must vary as in b), with three sign changes in
  $\phi''\equiv\partial_y^2\phi$ as in c).
\label{fig:proof} } 
\end{figure}

Consider firstly Eqns.~\ref{eqn:simplenavier} and~\ref{eqn:solveW}.
Sustituting $W_{xy}$ from Eqn.~\ref{eqn:solveW} into
Eqn.~\ref{eqn:simplenavier}, we obtain an expression for
$S(\gdot,\phi)$: this is just the family of homogeneous constitutive
curves, as plotted in Fig.~\ref{fig:spinodals} above.  Because the
constitutive equation is local, the solution at all points across the
rheometer cell must lie on one of these intrinsic constitutive curves.
Indeed, as the shear rate changes across the interface, the system
must pass through constitutive curves of differing concentrations to
maintain a uniform stress $\Sigma$. 
In other words, a relation $\phi=\phi(\gdot,\Sigma)$ must be
obeyed. The family of these curves is shown as dotted lines in
Fig.~\ref{fig:construction}a.
For the range of stresses at which $\phi(\gdot,\Sigma)$ is
non-monotonic, $\phi$ must have the form shown in
Fig.~\ref{fig:proof}b in which the derivative $\partial^2_y\phi$
changes sign three times across the interface, as in
Fig.~\ref{fig:proof}c.  [Actually, the forms of
Fig.~\ref{fig:proof}b,c assume that the profile in $\gdot$ increases
monotonically through the interface (Fig.~\ref{fig:proof}a). However
this monotonicity will emerge self consistently from our argument
below.]

However we know from Eqn.~\ref{eqn:third} that
$\partial_y^2\phi=\Lambda-\mu$.  $\Lambda$ is plotted in
Fig.~\ref{fig:construction}b,c using Eqn.~\ref{eqn:defineF} together
with the constraint $\phi=\phi(\gdot,\Sigma)$ (imposed from
Eqns.~\ref{eqn:simplenavier} and~\ref{eqn:solveW}, as discussed
above). From this plot we see that, for any $\mu$, a solution that
starts and ends in homogeneous phases (for which
$\partial_y^2\phi=\Lambda-\mu=0$)~\footnote{While this boundary
  condition is intuitively clear (the homogeneous phases are in
  general large compared with the interface), in our numerics we only
  actually imposed the boundary condtion $\partial_y\phi=0$.  We
  therefore did not, a priori, ``overspecify'' the differential
  equation~\ref{eqn:third}. However we shall show below that
  $\partial^2_y\phi$ must {\em automatically} equal zero at each
  interface.}  can only involve at most one sign change of
$\partial_y^2\phi$ between the boundaries.  This inconsistency with
Fig.~\ref{fig:proof}c means that a steady banded solution cannot exist
for these stress values for which $\phi(\gdot,\Sigma)$ is
non-monotonic. To summarize: for stresses far enough above the
critical point that $\phi(\gdot,\Sigma)$ is non-monotonic, a steady
state solution cannot simultaneously satisfy
Eqns.~\ref{eqn:simplenavier} and~\ref{eqn:solveW} (which imply three
sign changes of $\partial_y^2\phi$) at the same time as
Eqn.~\ref{eqn:third} (which only allows one sign change). Therefore
there a steady, smoothly banded profile cannot exist for such
stresses.

This argument is consistent with the sharp numerical profiles of
Fig.~\ref{fig:profiles_gammonly}, which are replotted in
Fig.~\ref{fig:construction}a (solid lines): each solution should have
followed a local (dotted) curve $\phi(\gdot)$, but instead has jumped
across the region in which this curve is non-monotonic. (The stresses
used to generate the homogeneous solutions $\phi(\gdot,\Sigma)$ in
Fig.~\ref{fig:construction}a were slightly different from those of the
numerical profiles: however the trend is still clear.) Although
Eqn.~(\ref{eqn:simpleconc}) is highly oversimplified, we believe that
the failure to negotiate the interface due to the conflict described
above is the reason for non-selection in the full, numerically solved
model. 

We return finally to justify our assumption that the shear rate must
increase monotonically through the interface, and to discuss in more
detail the nature of the banded solution when it {\em can} exist (\ie\ 
for stresses near crticial point where $\phi(\gdot)$ is monotonic).
Multiplying Eqn.~\ref{eqn:third} across by $d\phi/dy$, integrating on
$\phi$, and imposing $\partial_y\phi=0$ at each boundary, we find,
{\em for the simplified model of Eqn.~\ref{eqn:simpleconc}},
\begin{equation}
\label{eqn:equalareas}
\int_{\phi_l}^{\phi_r}d\phi\,\left[\mu-\Lambda\right]=0,
\end{equation}
which is an ``equal areas'' construction. ($\phi_l$ and $\phi_r$
denote the boundary values at $y=0,L$.) 
\begin{figure}[htb]
\includegraphics[scale=0.3]{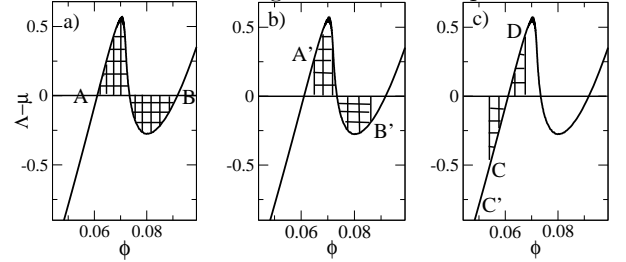}
\caption{Of these three proposed constructions specifying the banded
  state (when it is selected, near the critical point), only a) and b)
  are consistent with the boundary conditions $\partial_y\phi=0$. Of
  these, b) is for a finite system for which $\partial_y^2\phi\neq 0$
  at the boundary while a) is for the realistic physical limit in
  which the interface is narrow compared with the gap size,
  connecting two homogeneous phases in which
  $\partial_y^n\phi=0\,\forall\,n$. 
\label{fig:equalareas} } 
\end{figure}
If, in addition, we were to impose that $\partial_y^2\phi=0$ at each
boundary, then the construction must automatically be as shown in
Fig.~\ref{fig:equalareas}a. However we did not actually impose this
condition in our numerics, so the construction of
Fig.~\ref{fig:equalareas}b is also possible. This in fact corresponds
to a finite system, where the true homogeneous state
$\partial_y^n\phi=0\,\forall n$ is not quite reached at the
boundaries.  Any other equal areas construction
(Fig.~\ref{fig:equalareas}c) is not possible, for the following
reason.  Consider starting at point C with $\partial_y\phi=0$ (which
we {\em do} impose at the boundary in our numerics).
Eqn.~\ref{eqn:third} then tells us that $\partial_y^2\phi<0$ at this
point, so the function $\phi(y)$ must curve downwards from its
starting point of zero slope.  Therefore $\phi$ locally decreases, and
the system moves to point $C'$.  Repeating this argument, we find that
the system can never cross to the point $D$. By similar reasoning, the
shear rate must rise monotonically through the interface since any
initial fall (from the side of the low shear band) would be similarly
unstable to point $C$ in Fig.~\ref{fig:equalareas}c above.

Of course the concentration equation~(\ref{eqn:simpleconc}) is highly
oversimplified. For instance, a more realistic model (such as the one
of Eqn.~\ref{eqn:relative}) would have $\phi$ dependent prefactors to
the $\partial_y^2\phi$ term. {\em The equal areas result of
  Eqn.~(\ref{eqn:equalareas}) is therefore specific to our
  oversimplified Eqn.~(\ref{eqn:simpleconc}), and does not hold in
  general.}  Nonetheless we believe that Eqn.~\ref{eqn:simpleconc}
correctly predicts the absence of a uniquely banded solution for
stresses far above the critical point, via the basic conflict between
the number of sign changes of $\partial_y^2\phi$ across the interface,
described above.

\section{Conclusions}
\label{sec:conclusion}

In this paper, we have studied the role of concentration coupling in
the shear banding of complex fluids using the two-fluid, non-local
Johnson-Segalman model. We have calculated phase diagrams for
different degrees of coupling between concentration and mechanical
degrees of freedom (molecular strain), and found a phase diagram
qualitatively consistent with experiments on micellar solutions at
dilutions well below the equilibrium isotropic-to-nematic transition
\cite{berret94b}. Specific points to note are as follows.

\begin{enumerate}
  
\item The coexistence plateau in the steady-state flow curve slopes
  upward with shear rate, because of the concentration difference
  between the coexisting bands. The overall plateau height and width
  decrease with average concentration, terminating in a
  non-equilibrium critical point.  CPCl/NaSal in
  brine~\cite{berret94b} shows the same trend. 
  
\item 
  
  Of the two coexisting bands, the high shear band has a smaller
  concentration due to the fact that concentration tends to move up
  gradients in the normal micellar strain component $W_{yy}$ (where
  $y$ is the flow-gradient direction).  ($\tens{W}$ describes
  deformation relative to the unit tensor $\tens{\delta}$, and
  $W_{yy}$ is more negative in the high-shear phase than in the lower
  shear phase.) Tie lines of the phase diagram in the $\gdot,\phi$
  plane therefore have negative slope.

\item 
  The concentration gap is smaller for smaller values of
  concentration-coupling $\alpha\propto G'(\phi)/f''(\phi)$, and
  tends to zero in the limit $\alpha\to 0$. Accordingly, the
  coexistence region of the steady-state flow curve becomes flat in
  this limit.
  
\item We have described the way in which the flow phase diagram can be
  reconstructed from the family of flow curves $\Sigma(\gdotb,\phib)$,
  measured for several average concentrations $\phib$
  (Fig.~\ref{fig:reconstruct}).
  
\item The phase diagram and flow curves depend slightly on the
  relative size of the interfacial term in the viscoelastic
  constitutive equation to that in the equation that specifies the
  concentration dynamics.  This is a concrete demonstration of how
  stress selection and the coexistence conditions of driven systems
  depend on the nature of the interface, in contrast to equilibrium
  coexistence.
  
\item We find \textit{no} unique state selection when there are no
  gradient terms in the viscoelastic constitutive equation, except for
  stresses that are close to the critical point.  This implies that,
  for a model to reproduce a uniquely selected stress, it is not
  enough to simply have gradient terms only in, for example, the
  concentration dynamics. The dynamical equations of motion for each
  degree of freedom must possess inhomogeneous terms to attain
  selection in all situations. Conversely, in situations where such
  terms are physically absent, one can expect, under certain
  conditions, no selection and hence a range of control parameters
  (shear stress or strain rate) for which the steady states are
  \textit{intrinsically} history-dependent.
  
\item The interface width diverges at the critical point as a power
  law $(\Sigma-\Sigma_{\rm c})^{-n}$ with $n\approx 0.5$, although
  $n$ differs slightly across the different order parameters.

\end{enumerate}

Although our \model\ model is highly oversimplified, we believe that
it contains the basic ingredients required for a first description of
wormlike micellar surfactant solutions at concentrations well below
the isotropic-nematic (I-N) transition. In particular, it incorporates
the minimal set of realistic degrees of freedom (tensorial order
parameter for the micellar strain together with concentration), and
unifies a non-monotonic flow curve with the Helfand-Fredrickson
coupling between concentration and flow. Similar techniques could be
applied to more involved Cates non-linear theory for wormlike micelles
\cite{cates90,SpenCate94}.

We recall a previous calculations by Olmsted \etal\ was aimed at
systems of rigid rods near the I-N transition~\cite{olmsted99c}. In
future work we hope to unify these two approaches into a description
of wormlike micelles that is valid over the entire concentration
range. This should provide a first step towards understanding the
crossover regime in the data of Fig.~\ref{fig:coexistence}, in which the
coexistence plateau stress is a non monotonic function of the micellar
concentration.

\begin{acknowledgements}
  We thank J-F Berret and Paul Callaghan for useful discussions, and
  EPSRC GR/N11735 for financial support.
\end{acknowledgements}

\appendix
\bw

\section{ \model\ Equations in Cartesian Coordinates}
\label{app:fulleqns}

In this appendix, we give the components of the \model\ model's
equations for planar shear flow along the $x$ direction, allowing
gradients only in the flow-gradient direction, $y$, as described in
Sec.~\ref{sec:geometry}, above. The $x$ component of force-balance is
(in the zero-Reynolds limit considered in this paper)
\be 
0=\partial_y
\left[ G \left( \phi \right) { W_{xy}} \right] +\,{ \etam}\,\partial_y
\left[\,\phi\,\partial_y { v_{mx}} \right] +\,{ \etas}\,\partial_y
\left[ \, \left( 1-\phi \right) \partial_y { v_{sx}} \right].  \ee 
The $y$ component of force-balance is fixed by incompressibility,
$\nablu.\vect{v}=0$, along with the boundary condition $v_y=0$:
\be
0=\phi v_{my}+(1-\phi)v_{sy}.
\ee
The relative velocity between the micelles and solvent (again ignoring
inertial terms) is
\begin{align}
v_{my}-{ v_{sy}}&=\frac{\phi(1-\phi)}{\zeta} \left\{ \frac{1}{\phi}
  \partial_y \left[ G \left( \phi 
 \right) { W_{yy}} \right] +2\,\frac{1}{\phi}  { \etam}
\,\partial_y \left[  \,\phi\,\partial_y  { v_{my}}   \right] -2\, \frac{1}{1-\phi} { \etas}\,\partial_y \left[ \, \left( 1-\phi
 \right) \partial_y  { v_{sy}}   \right] - \partial_y \mathfrak{F} \right\}\\
 { v_{mx}}-{ v_{sx}}&=\frac{\phi(1-\phi)}{\zeta}\left\{ \frac{1}{\phi}
   \partial_y \left[ G \left( \phi 
 \right) { W_{xy}} \right] +\,\frac{1}{\phi}  { \etam}
\,\partial_y \left[  \,\phi\,\partial_y  { v_{mx}}   \right] - \,\frac{1}{1-\phi} { \etas}\,\partial_y \left[ \, \left( 1-\phi
 \right) \partial_y { v_{sx}}  \right]\right\}\\
\mathfrak{F}&= { f'} \left( \phi \right) - \,g  \partial^2_y \phi  + \tfrac{1}{2}\,{ G'}
 \left( \phi \right)  \left[ { W_{yy}}+{ W_{xx}}-\ln  \left( { W_{yy}}
\,{ W_{xx}}+{ W_{yy}}+{ W_{xx}}+1-{{ W_{xy}}}^{2} \right)  \right].
\end{align}
The evolution of the micellar strain tensor is given by
\begin{align}
\partial_t W_{xy}+{ v_{my}}\,\partial_y  { W_{xy}}&=  \tfrac{1}{2}(a-1) W_{xx} \,\partial_y  { 
v_{mx}} + \tfrac{1}{2}(1+a)\,W_{yy} \partial_y  { v_{mx}}  +a
  \,W_{xy} \partial_y  { v_{my}} + \,\partial_y  { 
v_{mx}}   
-{\frac {{ W_{xy}}}{\tau \left( \phi \right) }}+{\frac {
 l \left( \phi \right) ^{2} \partial^2_y { W_{xy}}
  }{\tau \left( \phi \right) }},\\
\partial_t W_{yy}+{ v_{my}}\,\partial_y  { W_{yy}} &=  (a-1)
\,W_{xy}\partial_y  {  
v_{mx}} +2\,a   \,W_{yy} \partial_y  { v_{my}} + 2\,\partial_y  { v_{my}}   -{\frac {{
 W_{yy}}}{\tau \left( \phi \right) }}+{\frac {  l \left( \phi
 \right)   ^{2} \partial^2_y  { W_{yy}}  }{
\tau \left( \phi \right) }},\\
\partial_t W_{xx}+{ v_{my}}\,\partial_y { W_{xx}} &=  (1+a)\,W_{xy}\partial_y  { 
v_{mx}}  -{
\frac {{ W_{xx}}}{\tau \left( \phi \right) }}+{\frac { l
 \left( \phi \right)   ^{2}\partial^2_y  { W_{xx}} 
 }{\tau \left( \phi \right) }}.
\end{align}
Finally, the concentration dynamics are
\bea 
\partial_t\phi&=&-\partial_y \left\{ \frac{\phi^2\, \left( 1-\phi
    \right)^2}{\phi\zeta}
  \left\{ \partial_y \left[ G \left( \phi
      \right) { W_{yy}} \right] +2\,\etam \,\partial_y \left[
      \,\phi\,\partial_y  { v_{my}} 
       \right] -\frac{2\phi}{1-\phi}  {
      \etas}\,\partial_y \left[ \, \left( 1-\phi \right) \partial_y
      { v_{sy}}\right] - \partial_y \mathfrak{F} \right\} \right\}.  \eea

\ew

\section{Stationary homogeneous  solutions of the \model\ model}
\label{app:homogen}
  
In planar shear, the stationary homogeneous solutions to
Eqns.~(\ref{eqn:navier}-\ref{eqn:JSd}) for given $\gdot$ and $\phi$
are $\vrel\equiv\vm-\vs=0$ and
\begin{subequations}
\label{eqn:intrinsic_flow_curve}
\begin{align}
W_{xy}&=\frac{\gdot\tau(\phi)}{1+b\gdot^2\tau^2(\phi)},\\
W_{yy}&= \frac{a-1}{1+a}W_{xx}=
-\frac{1}{(1+a)}\frac{b\gdot^2}{1+b\gdot^2}\\
W_{zz}&=W_{xz}=W_{yz}=0,
\end{align}

\end{subequations}
where $b=1-a^2$. The steady state shear stress is given by 
\begin{equation}
  \label{eq:3}
  \Sigma_{xy}=G(\phib)W_{xy}+\phi\etam+(1-\phi)\etas\gdot=\textrm{constant}.
\end{equation}



\end{document}